\documentclass[prb,aps,twocolumn,amsmath,amssymb,floatfix,
superscriptaddress]{revtex4}
\usepackage[dvips]{graphics}
\usepackage[justification=raggedright]{caption}
\usepackage[title]{appendix}
\usepackage{subcaption}
\usepackage{color}
\definecolor{dred}{rgb}{0,0,0.6}
\usepackage{graphicx}    
\usepackage{bm}        
\usepackage{amssymb}   
\usepackage{amsmath}
\usepackage{braket}
\usepackage[mathscr]{euscript}
\usepackage[colorlinks=true, allcolors=blue]{hyperref}
\begin{document}
	
	\title{Properties of the non-Hermitian SSH model: role of $\mathcal{PT}$ symmetry}
	
	\author{Dipendu Halder}
	\affiliation{Department of Physics, Indian Institute of Technology Guwahati-Guwahati, 781039 Assam, India}
	\author{Sudin Ganguly}
	\affiliation{Department of Physics, School of Applied Sciences, University of Science and Technology Meghalaya, Ri-Bhoi-793101, India}
	\author{Saurabh Basu}
	\affiliation{Department of Physics, Indian Institute of Technology Guwahati-Guwahati, 781039 Assam, India}
	\begin{abstract}
		The present work addresses the distinction between the topological properties of $\mathcal{PT}$ symmetric and non-$\mathcal{PT}$ symmetric scenarios for the non-Hermitian Su-Schrieffer-Heeger (SSH) model.
		The non-$\mathcal{PT}$ symmetric case is represented by non-reciprocity in both the inter- and the intra-cell hopping amplitudes, while the one with $\mathcal{PT}$ symmetry is modeled by a complex on-site staggered potential.
		In particular, we study the loci of the exceptional points, the winding numbers, band structures, and explore the breakdown of bulk-boundary correspondence (BBC).
		We further study the interplay of the dimerization strengths on the observables for these cases.
		The non-$\mathcal{PT}$ symmetric case denotes a more familiar situation, where the winding number abruptly changes by half-integer through tuning of the non-reciprocity parameters, and demonstrates a complete breakdown of BBC, thereby showing non-Hermitian skin effect.
		The topological nature of the $\mathcal{PT}$ symmetric case appears to follow closely to its Hermitian analogue, except that it shows unbroken (broken) regions with complex (purely real) energy spectra, while another variant of the winding number exhibits a continuous behavior as a function of the strength of the potential, while the conventional BBC is preserved.
	\end{abstract}
	\maketitle

	\section{\label{sec1}Introduction}
	In conventional quantum mechanics, the Hamiltonian of a system is Hermitian, which implies that the dynamical observables assume real values. 
	Thus isolated systems are dealt within Hermitian quantum mechanics, while real systems always interact with the environment, and then dissipate. 
	Hence an exchange of energy between the system under consideration and the `bath' occurs.
	These are called open systems, and in a real scenario, the coupling to the environment makes their description significantly more complex.
	Strictly speaking, the time evolution of the wave function via a Hermitian Hamiltonian needs to be replaced by a Liouvillian superoperator via the time evolution of the density matrices.\cite{Breuer}
	There are more technically challenging techniques to deal with open systems, such as, Keldysh formalism\cite{Keldysh}, Lindblad master equation\cite{Lindblad} etc.
	Simpler reconciliations in the form of using non-Hermitian (NH) Hamiltonians yield an intuitive understanding of the emergent phenomena in real physical systems, such as, photonic lattices with gain and/or loss, other optical systems,\cite{Eichelkraut,Li,Kremer,Yang,PhysRevLett.123.213903,PhysRevLett.115.040402,Xu,Xiao,Babak,Miguel,PhysRevLett.123.230401,Wang,Weimann,Goblot,PhysRevLett.120.113901,Pan} electronic,\cite{PhysRevB.97.220301,PhysRevLett.123.193901} and mechanical\cite{Brandenbourger,Ananya} systems etc.
	\par Since the introductory findings by Hatano and Nelson\cite{PhysRevLett.77.570,PhysRevB.56.8651} and Bender,\cite{PhysRevLett.80.5243} claiming that NH systems with certain symmetry, such as a combination of both the parity ($\mathcal{P}$) and the time reversal symmetry ($\mathcal{T}$), namely the $\mathcal{PT}$ symmetry, can have real energy spectra, they have emerged as topics that are worth exploring.
	In the context of present day's research, $\mathcal{PT}$ symmetric Hamiltonians have emerged as appropriate description of the dissipative systems with balanced gain and loss,\cite{Ganainy} and thus have been attracting a lot of attention.
	In certain cases, the Hermitian Hamiltonians reside at the borderline of $\mathcal{PT}$ symmetric and non-$\mathcal{PT}$ symmetric cases. 
	In our work, we shall consider two different NH Hamiltonians, one $\mathcal{PT}$ symmetric, while $\mathcal{PT}$ symmetry is absent for the other.
	\par Meanwhile, a good volume of research has also concentrated on topological band theory extended to NH systems, where the topological classification gets significantly richer than their Hermitian counterpart\cite{PhysRevX.9.041015}. 
	Besides, in a Hermitian system, the topological phases are characterized by gapless edge modes in the open boundary condition (OBC).
	However, no such edge modes exist under the periodic boundary condition (PBC), as the PBC corresponds to an infinite system.
	Yet some topological invariants may be used to garner information of the topological phase transition in finite systems (OBC). 
	This correspondence is known as the bulk boundary correspondence (BBC), which is no longer valid in NH systems (or have to be modified), thereby constituting an important deviation from its Hermitian counterpart, and the phenomena is termed as the breakdown of BBC.\cite{PhysRevLett.102.065703,PhysRevB.84.153101,PhysRevB.84.205128,PhysRevA.89.062102,PhysRevLett.116.133903,PhysRevLett.118.040401,PhysRevLett.120.146402,PhysRevB.97.045106,PhysRevA.97.052115,PhysRevLett.121.086803,PhysRevLett.121.026808,PhysRevB.97.121401,PhysRevX.8.031079,PhysRevLett.121.136802,PhysRevB.99.081103,PhysRevB.98.165148,PhysRevB.98.094307,PhysRevA.98.052116,PhysRevLett.123.246801,PhysRevLett.124.056802}
	Moreover, systems with OBC and PBC are way too far distinct from one another in NH systems, which is not the case for the Hermitian analogues.
	Another fascinating manifestation is the non-Hermitian skin effect (NHSE), which is related to localization property at the edges of the system.
	In several recent works,\cite{PhysRevLett.121.026808,PhysRevLett.121.086803,PhysRevX.8.031079} it has been shown that a macroscopic number of eigenstates get localized at either edge of the system, as soon as non-Hermiticity is introduced.
	This may be considered as a direct consequence of the breakdown of BBC.
	\par In addition to these, the existence of exceptional points (EP),\cite{Heiss_2012,Hsu,PhysRevX.6.021007,PhysRevB.101.045130,PhysRevA.101.033820,PhysRevLett.102.065703} where the Hamiltonian becomes ill-defined owing to formation of Jordan blocks, and hence the linearly independent vectors fall short of the degeneracy of the eigenvalues.
	Simply speaking, EPs are the singularities in the system where all the eigenvalues and the eigenvectors of a system coalesce, making the Hamiltonian non-diagonalizable.
	They play a major role in segregating the topologically trivial, and the non-trivial phases of a NH system.
	Plenty of theoretical works have suggested that the NH systems can have non-trivial properties.
	Implementation of NH physics in systems with distinct topological characteristics, such as, models like Su-Schrieffer-Heeger (SSH) models\cite{PhysRevB.97.045106,PhysRevB.84.153101,PhysRevLett.123.066404,Schomerus,PhysRevA.89.062102,YUCE20151213,PhysRevB.95.174506,PhysRevA.101.013635,PhysRevB.103.014302,PhysRevA.101.063839,PhysRevB.103.125411,Shuai,Cui:20,PhysRevA.100.012112,PhysRevA.100.032102,PhysRevB.102.041119,PhysRevB.100.165430,Poco}, Aubry-André-Harper models\cite{PhysRevB.103.054203,PhysRevB.100.125157,PhysRevA.103.L011302,PhysRevB.101.020201,PhysRevB.101.235150,PhysRevResearch.2.033052}, Rice-Mele models\cite{PhysRevA.98.042120,PhysRevLett.49.1455,PhysRevA.99.032109,PhysRevB.48.4442,PhysRevLett.125.186802} etc. have been widely studied to ascertain the interplay of topology and non-Hermiticity.
	\par In this paper, we have taken the simplest system that encodes topological considerations, such as a dimerized tight binding model in one dimension, or the familiar SSH model, which is thought to be a realistic model for polyacetylene in order to demonstrate the distinction between the $\mathcal{PT}$ symmetric or the non-$\mathcal{PT}$ symmetric systems.
	A main feature of the SSH model is the existence of two topologically different phases that are distinguished by the presence or the absence of two-fold degenerate zero-mode edge states (one at each edge) under the OBC.
	The topological invariant that provides information on these zero-mode edge states is the winding number\cite{RevModPhys.82.3045,RevModPhys.83.1057,Bernevig+2013,Asboth}, which, in Hermitian systems, can take only integral values.
	In the conventional Hermitian SSH model, the value of winding number is zero and unity for the topologically trivial and the non-trivial regimes respectively.
	\par In our model, non-Hermiticity is introduced in two different ways; first through a non-reciprocity in the hopping integrals (both within and across the unit cells), and second, introducing a complex staggered on-site potential in the SSH model. 
	The two resultant Hamiltonians differ with regard to their $\mathcal{PT}$ symmetry, with the former denoting a case with absent  $\mathcal{PT}$ symmetry, while the latter preserves $\mathcal{PT}$ symmetry.
	We ascertain different topological properties of the two models. 
	Specifically, we compare and contrast between the corresponding topological phases, NHSE, the structure of the EPs, and the winding numbers.
	Further, we investigate the interplay of the dimerization strength (ratio of the intra-cell and the inter-cell hopping amplitudes) with these observables.
	\par Our paper is organized as follows.
	We introduce the Hamiltonians corresponding to the $\mathcal{PT}$ symmetric, and the non-$\mathcal{PT}$ symmetric cases in section II.
	Hence we discuss the band structure, NHSE, and the role of the corresponding EPs in section III.
	Further, we compute the winding numbers in each of these cases to ascertain their topological characteristics.
	In fact, another distinct definition of the winding number for the $\mathcal{PT}$ symmetric case is shown to have a continuous variation as a function of the strength of the on-site potential, implying that it can assume any value in the range $[0:1]$.
	Further, the non-$\mathcal{PT}$ symmetric case demonstrates NHSE, which is absent in the $\mathcal{PT}$ symmetric case.
	Finally, we conclude with a brief summary of our results in section IV.
	
	\section{\label{sec2}Model Hamiltonians}
	
\begin{figure}[ht]
		\includegraphics[width=0.5\textwidth]{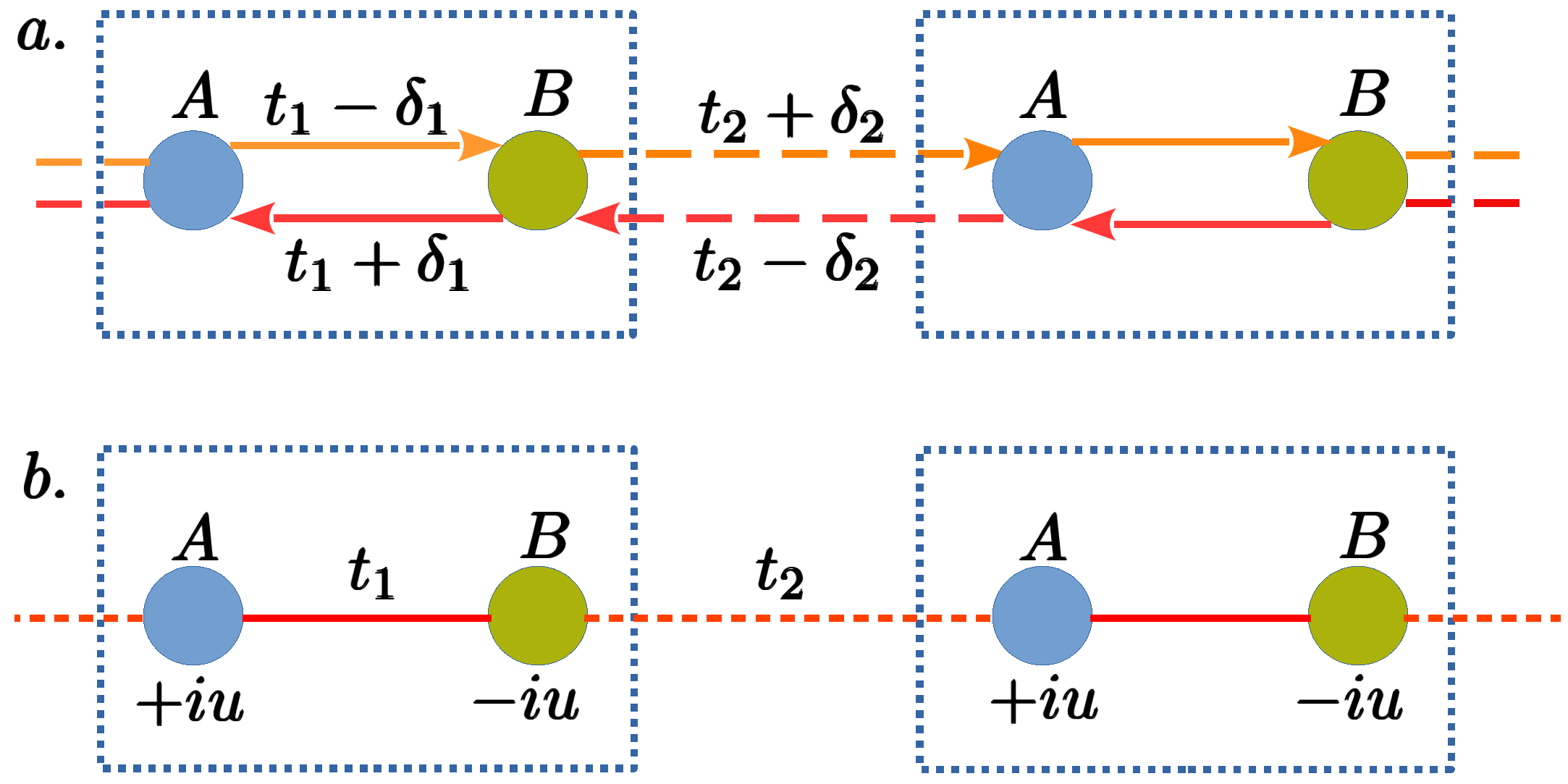}
		\caption{(a) NH SSH model with non-hermiticity induced in the intra- ($t_1$) and the inter-cell ($t_2$) hopping amplitudes. (b) NH SSH model with imaginary potential, $u$. The box represents a single unit cell.}
		\label{Fig:NH-Systems}
	\end{figure}

	We have incorporated the non-Hermiticity in the tight binding model with two atoms per unit cell, that is, the so called SSH model in the following fashion.
	First, we consider a model with non-reciprocity in both the intra-cell, and the inter-cell hopping energies, which are pictorially shown in Fig.\ref{Fig:NH-Systems}(a).
	The corresponding Hamiltonian can be written as,
	\begin{align}
		\hat{H}_1=\sum_n\bigg[&(t_1-\delta_1)\hat{a}_n^{\dagger}\hat{b}_n+(t_1+\delta_1)\hat{b}_n^{\dagger}\hat{a}_n+\nonumber\\
		&(t_2-\delta_2)\hat{b}_n^{\dagger}\hat{a}_{n+1}+(t_2+\delta_2)\hat{a}_{n+1}^{\dagger}\hat{b}_n\bigg]\label{eq:Ham1}
	\end{align}
	where, $\hat{a}_n$($\hat{a}_n^{\dagger}$) and $\hat{b}_n$($\hat{b}_n^{\dagger}$) are annihilation(creation) operators, corresponding to the sites at the A and the B sublattices, respectively, of the $n^{th}$ unit cell, and $t_1$($t_2$) is intra-cell(inter-cell) hopping amplitude with a non-reciprocity in $\delta_1$($\delta_2$).
	An alternate NH model may be obtained by including a staggered imaginary on-site potential, while the hopping parameters are kept unaltered.
	This Hamiltonian can be written as,
	\begin{align}
		\hat{H}_2&=\sum_n\bigg[iu\left(\hat{a}_n^{\dagger}\hat{a}_n-\hat{b}_n^{\dagger}\hat{b}_n\right)+\nonumber\\
		&t_1(\hat{a}_n^{\dagger}\hat{b}_n+\hat{b}_n^{\dagger}\hat{a}_n)+
		t_2(\hat{b}_n^{\dagger}\hat{a}_{n+1}+\hat{a}_{n+1}^{\dagger}\hat{b}_n)\bigg]\label{eq:Ham2}
	\end{align}
	where $u$ is the strength of the imaginary on-site potential.
	The system can be visualized as presented in Fig.\ref{Fig:NH-Systems}(b).
	\par To make the problem more tractable, we Fourier transform $\hat{H}_j$ ($j \in 1,2$) and write them in the Bloch form as,
	\begin{gather}
	{H}_j=\sum_k\begin{pmatrix}
	\hat{a}_k^{\dagger} & \hat{b}_k^{\dagger}
	\end{pmatrix}h_j(k)
	\begin{pmatrix}
	\hat{a}_k\\
	\hat{b}_k
	\end{pmatrix}
	\end{gather}
	where,
	\begin{gather}
		h_1(k)=\begin{pmatrix}
		0 & (t_1-\delta_1)+(t_2+\delta_2)e^{-ik}\\
		(t_1+\delta_1)+(t_2-\delta_2)e^{ik} & 0
		\end{pmatrix}
		\label{eq:mat}
	\end{gather}
	and
	\begin{gather}
	h_2(k)=\begin{pmatrix}
	iu & t_1+t_2e^{-ik}\\
	t_1+t_2e^{ik} & -iu
	\end{pmatrix}.
	\label{eq:u}
	\end{gather}
	Both Eqs.\eqref{eq:mat} and \eqref{eq:u} can be expressed in the form of the standard Dirac Hamiltonian, namely,
	\begin{equation}
	H_j=\pmb{d}_j\cdot\pmb{\sigma}
	\end{equation}
	where the $\pmb{d}_j$ denote vectors in the complex plane.
	Further the components of the $\pmb{d}_j$-vector are written as,
	\begin{eqnarray}
		\pmb{d}_{1}^\mathrm{R}(k) & = & \big( t_1+t_2\cos{k},\;t_2\sin{k},\;0 \big) \nonumber\\
		\pmb{d}_{1}^\mathrm{I}(k) & = & \big(-\delta_2\sin{k},\;-\delta_1+\delta_2\cos{k},\;0\big)
		\label{eq:d}
	\end{eqnarray}
	and
	\begin{eqnarray}
		\pmb{d}_{2}^\mathrm{R}(k) & = & \big( t_1+t_2\cos{k},\; t_2\sin{k},\; 0 \big) \label{eq:du}\\
		\pmb{d}_{2}^\mathrm{I} & = & \big(0,\;0,\;u\big)\nonumber
	\end{eqnarray}
	where, $\pmb{\sigma}$ denote the Pauli matrices, and $\pmb{d}_{j}^\mathrm{R}$, $\pmb{d}_{j}^\mathrm{I}$ represent the real and the imaginary parts of $\pmb{d}_j$ respectively.
	The eigenvalues of these Hamiltonians are given by,
	\begin{equation}
		E_j=\pm|\pmb{d}_j|.
		\label{eq:eig}
	\end{equation}
	Let us examine the symmetries of the Hamiltonians, $h_j$.
	$h_1(k)$ clearly has a chiral symmetry, which is evident from the following relation,\cite{PhysRevB.97.045106}
	\begin{equation}
		\sigma_z\ h_1(k)\ \sigma_z=-h_1(k)
		\label{eq:c}
	\end{equation}
	However, it does not possess the $\mathcal{PT}$ symmetry, that is,
	\begin{equation}
		\sigma_x\ h_1(k)\ \sigma_x\ne h_1^*(k)
	\end{equation}
	While the opposite happens for $h_2(k)$, that is, the $\mathcal{PT}$ symmetry exists,
	\begin{equation}
		\sigma_x\ h_2(k)\ \sigma_x=h_2^*(k)
		\label{eq:pt}
	\end{equation}
	but the chiral symmetry is lost, namely,
	\begin{equation}
		\sigma_z\ h_2(k)\ \sigma_z\ne -h_2(k).
	\end{equation}
	\section{Results}
	In order to make the preceding discussion more structured, in the following section, we segregate the discussion of our results for the cases corresponding to the one with absent $\mathcal{PT}$ symmetry and the one with $\mathcal{PT}$ symmetry. 
	Specifically, we show the calculation of EPs, winding numbers, energy spectra, and demonstrate NHSE via the sensitivity (or its lack thereof) to the boundary conditions in both these situations. 
	While doing so, we have considered two cases that correspond to the topological and the trivial phases for the Hermitian SSH model, namely, $t_1<t_2$ and $t_1>t_2$ respectively.
	While we agree that the trivial and the topological phases of the Hermitian model might loose their significance in the NH analogue, we still use these regimes as the benchmarks to ascertain the role of the dimerization strength on the properties of our NH models.
	Specifically, all the while we consider $t_{1} =1$, and fix $t_{2}$ at $2$ and $0.5$ corresponding to the topological and the trivial regimes, respectively.
	
	\subsection{non-$\mathcal{PT}$ symmetric case}
	In the non-$\mathcal{PT}$ symmetric case, the components of the $\pmb{d}$-vector can be expressed via a complex angle $\phi$, which is defined by,
	\begin{equation}
	\tan{\phi}=\frac{d_y}{d_x}=\frac{d_{y}^\mathrm{R}+id_{y}^\mathrm{I}}{d_{x}^\mathrm{R}+id_{x}^\mathrm{I}}.
	\label{eq:tan}
	\end{equation}
	$\phi$ can be termed as an angle between the components ($d_{x},d_{y}$), and can be expressed as,
	\begin{equation}
	\phi=\phi_\mathrm{R}+i\phi_\mathrm{I}
	\label{eq:sum}
	\end{equation}
	where $\phi_\mathrm{R}$ and $\phi_\mathrm{I}$ denote the real and the imaginary parts of $\phi$. These components can further be expressed in terms of the components of the $\pmb{d}$-vector (see Eq.\eqref{eq:tan}),\cite{PhysRevA.97.052115}
	\begin{equation}
	e^{2i\phi_\mathrm{R}}=\frac{d_+}{d_-}\bigg/\left|\frac{d_+}{d_-}\right|,\quad\mathrm{and}\quad e^{-2\phi_\mathrm{I}}=\left|\frac{d_+}{d_-}\right|.
	\label{eq:r}
	\end{equation}
	Further, it can be shown that,
	\begin{equation}
	\tan 2\phi_\mathrm{R}=\mathrm{Im}\left(\frac{d_+}{d_-}\right) \bigg/ \mathrm{Re}\left(\frac{d_+}{d_-}\right)
	\label{eq:phi}
	\end{equation}
	where $d_{\pm}$ are defined by,
	\begin{align}
	d_+ &=d_{x}^\mathrm{R}-d_{y}^\mathrm{I}+i(d_{y}^\mathrm{R}+d_{x}^\mathrm{I})\quad\\ d_- &=d_{x}^\mathrm{R}+d_{y}^\mathrm{I}-i(d_{y}^\mathrm{R}-d_{x}^\mathrm{I})
	\end{align}
	$d_{x}^\mathrm{R}=\mathrm{Re}(d_x)$, $d_{x}^\mathrm{I}=\mathrm{Im}(d_x)$, $d_{y}^\mathrm{R}=\mathrm{Re}(d_y)$ and $d_{y}^\mathrm{I}=\mathrm{Im}(d_y)$.
	\begin{figure}[!h]
		\includegraphics[width=0.5\textwidth]{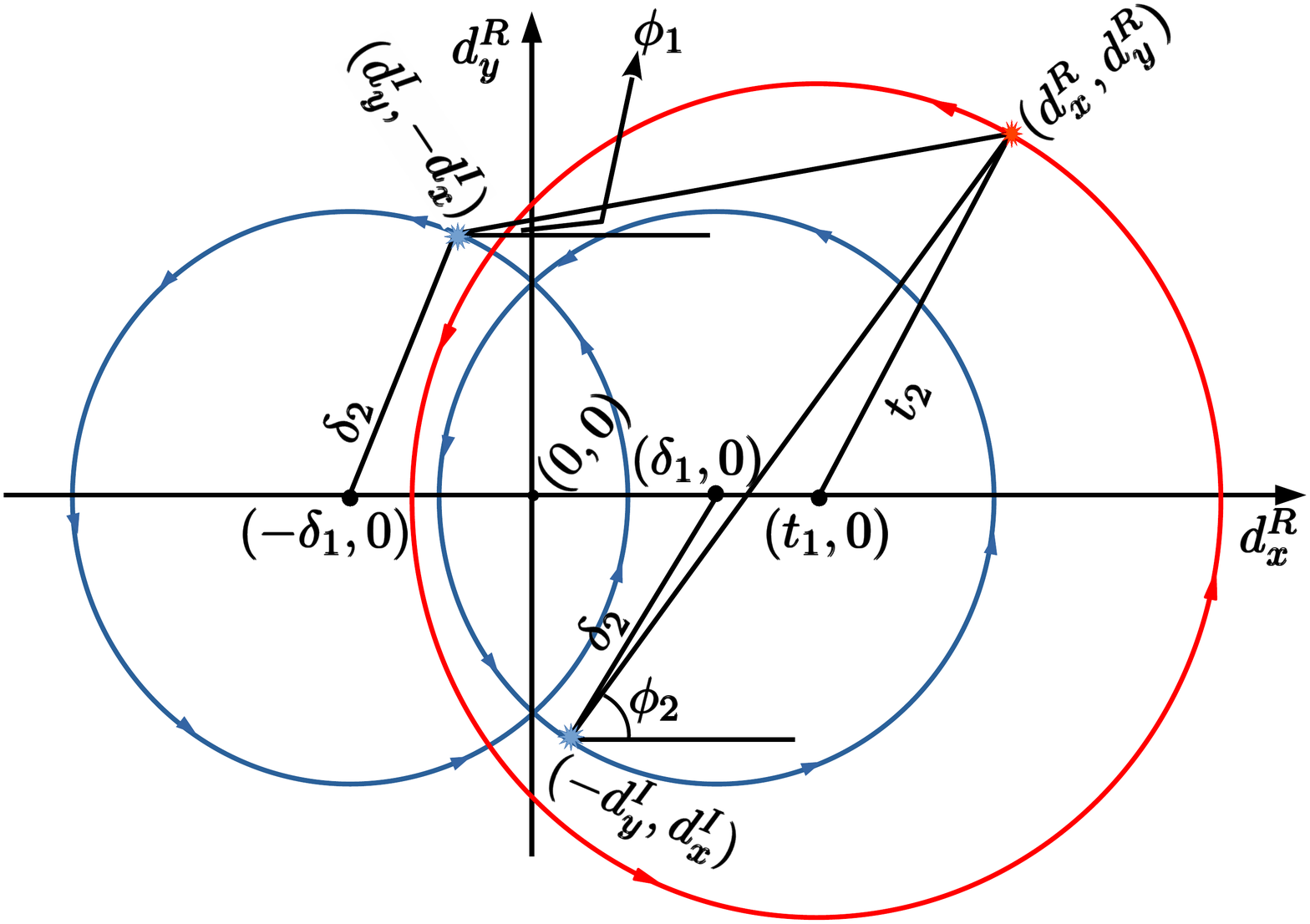}
		\caption{(Color online) Loci of the EPs and ($d_{x}^\mathrm{R}$, $d_{y}^\mathrm{R}$) in a space spanned by $d_{x}^\mathrm{R}$ and $d_{y}^\mathrm{R}$. The two blue circles are the loci of the EPs $(d_{y}^\mathrm{I}, -d_{x}^\mathrm{I})$ and $(-d_{y}^\mathrm{I}, d_{x}^\mathrm{I})$ with the blue stars representing the instantaneous position of the EPs for some non-zero values of $k$. The red star on the red circle represents instantaneous position of ($d_{x}^\mathrm{R}$, $d_{y}^\mathrm{R}$).}
		\label{Fig:Locus-EP}
	\end{figure}
	\par Let us focus on finding the EPs of this system.
	The eigenvalues given by Eq.\eqref{eq:eig} will coalesce when $E_{\pm}=0$, that is, when the following condition is satisfied,
	\begin{equation}
	(d_{x}^\mathrm{R}+id_{x}^\mathrm{R})^2+(d_{y}^\mathrm{R}+id_{y}^\mathrm{I})^2=0.
	\label{eq:e}
	\end{equation}
	From the above equation (Eq.\eqref{eq:e}), it is evident that the real part of the energy will be zero when
	\begin{equation}\label{eq:ep}
	d_{x}^\mathrm{R}=\pm d_{y}^\mathrm{I}\qquad\mathrm{and}\qquad d_{y}^\mathrm{R}=\mp d_{x}^\mathrm{I}.
	\end{equation}
	Further, the imaginary part of Eq.\eqref{eq:e} becomes zero when the hopping amplitudes satisfy,
	\begin{equation}
	\frac{t_1}{t_2}=-\frac{\delta_1}{\delta_2}.
	\end{equation}
	\begin{figure}[h!]
		\includegraphics[width=0.5\textwidth]{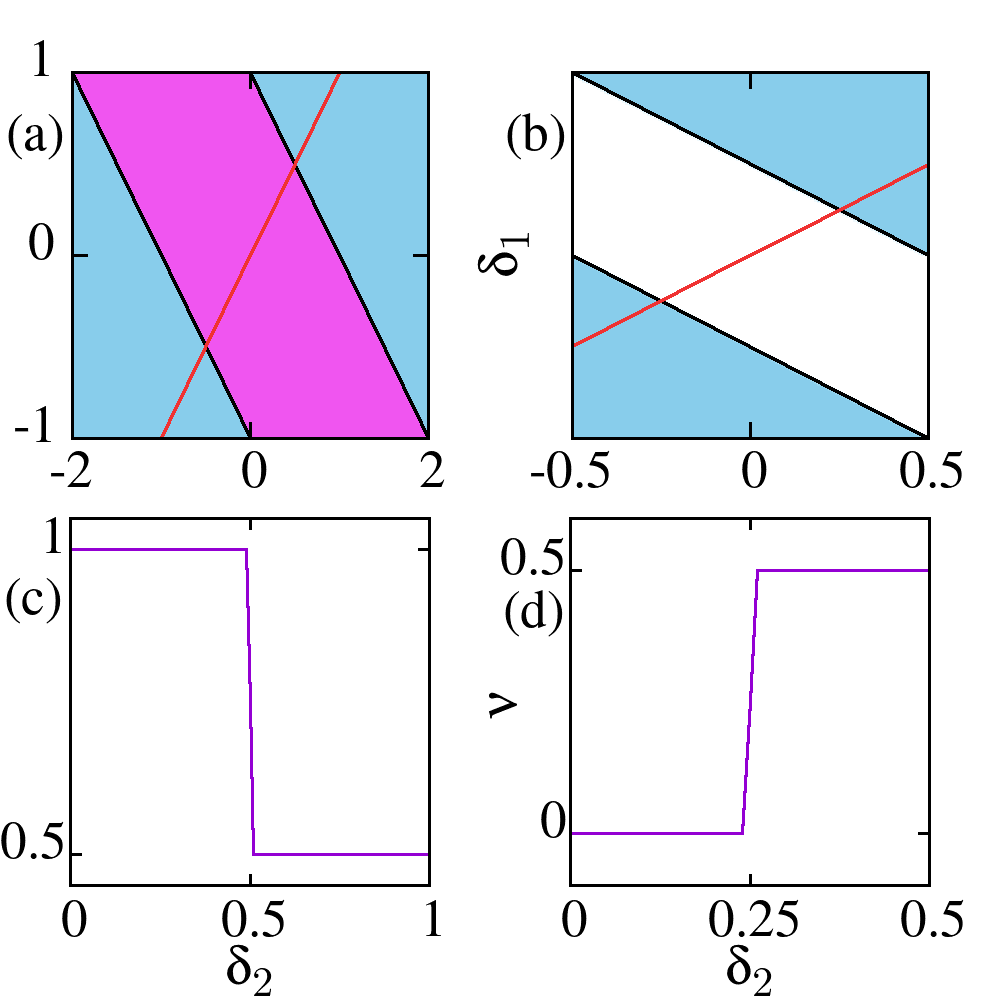}
		\caption{(Color online) Phase diagram of the winding number (0 `white’, 0.5 `sky blue’, 1 `light-magenta’) as a function of $\delta_1$ and $\delta_2$ for (a) $t_1=1$, $t_2=2$ (b) $t_1=1$, $t_2=0.5$. The discontinuous jumps in $\nu$ for (c) $t_1=1$, $t_2=2$ (d) $t_1=1$, $t_2=0.5$ are shown in the figures in lower panel.}
		\label{Fig:PhaseDiagrams-nu}
	\end{figure}
	\par Now, in an NH system, the left eigenvector (eigenvector of $H^{\dagger}$), and the right eigenvector (eigenvector of $H$) differ from each other, as the eigenvectors no longer form an orthonormal set due to the non-Hermiticity.
	Instead they satisfy the bi-orthonormal condition given by,
	\begin{equation*}
		\left<\Psi^\mathrm{{LE}}_n|\Psi^\mathrm{{RE}}_m\right>=\delta_{nm}
	\end{equation*}
	where $\Psi^\mathrm{{LE}}_n$ and $\Psi^\mathrm{{RE}}_m$ are the left and right eigenvectors corresponding to the eigenvalues $E_n^*$ and $E_m$ respectively.
	\par The right eigenvector of the Hamiltonian, $h_1(k)$, is given by,
	\begin{gather}
	\left|\Psi^\mathrm{{RE}}_{1\pm}\right>=\sqrt{\frac{R_1}{R_1+R_2}}\;e^{i\xi}\;\begin{pmatrix}
	\pm\sqrt{\frac{R_1}{R_2}}\;e^{-i(\alpha_1+\alpha_2)/2}\\
	1
	\end{pmatrix}
	\end{gather}
	with $ d_{-}=R_1e^{i\alpha_1}$ and $d_{+}=R_2e^{i\alpha_2}$, $e^{i\xi}$ being a phase factor.
	Further, $R_1=|(t_1-\delta_1)+(t_2+\delta_2)e^{ik}|$, $R_2=|(t_1+\delta_1)+(t_2-\delta_2)e^{-ik}|$, $\alpha_1=\tan^{-1}\left[\frac{(t_2+\delta_2)\sin k}{(t_1-\delta_1)+(t_2+\delta_2)\cos k}\right]$ and $\alpha_2=\tan^{-1}\left[\frac{(t_2-\delta_2)\sin k}{(t_1+\delta_1)+(t_2-\delta_2)\cos k}\right]$.
	Now the coalescence of the eigenvectors at the EPs demands that $R_1$ has to vanish.
	This is equivalent to the condition,
	\begin{equation}\label{eq:ep2}
	\big|\delta_1-\delta_2\big|=t_1+t_2\quad\mathrm{and}\quad\big|\delta_1+\delta_2\big|=\big|t_1-t_2\big|
	\end{equation}
	at $k=\pm\pi$ and $k=0$ respectively.
	\begin{figure}[ht]
		\includegraphics[width=0.45\textwidth]{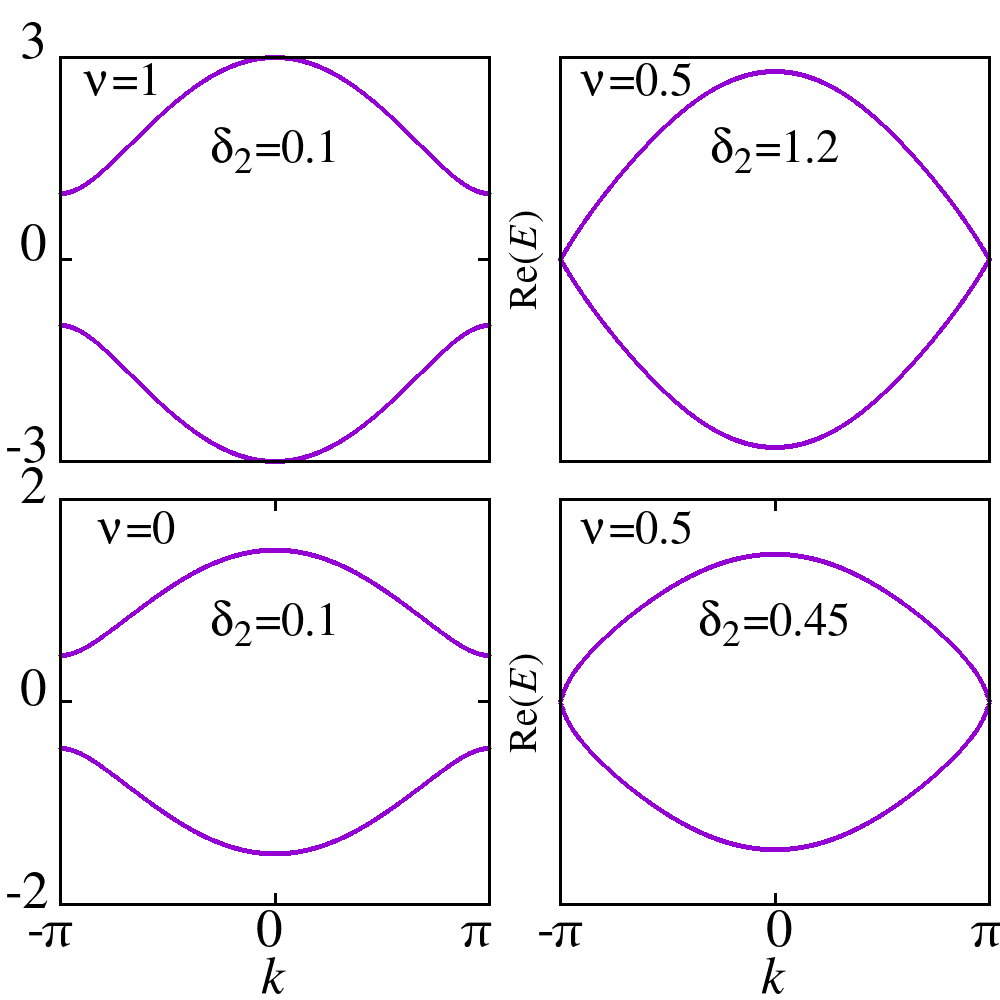}
		\caption{(Color online) Band structure (Re($E$) as a function of $k$) for $\delta_1=0.1\ \mathrm{and}\ t_2=2$ (first row) with $\delta_{2}=0.1\;\mathrm{and}\; 1.2$, and $\delta_1=0.3 \ \mathrm{and}\  t_2=0.5$ (second row) with $\delta_{2}=0.1\;\mathrm{and}\; 0.45$. We have kept $t_1=1$. The spectral gaps close in right panel.}
		\label{Fig:b1}
	\end{figure}
	\par Hence the location of the EPs is given by Eq.\eqref{eq:ep} along with the criteria laid down in Eq.\eqref{eq:ep2}.
	Now, the real part of the angle $\phi$ (see Eq.\eqref{eq:phi}) can be written in terms of the location of the EPs, that is,
	\begin{equation}
	\tan 2\phi_\mathrm{R}=\frac{\tan \phi_1+\tan \phi_2}{1-\tan \phi_1\tan \phi_2}=\tan (\phi_1+\phi_2)
	\end{equation}
	where the angles $\phi_{1}$ and $\phi_{2}$ are given by,
	\begin{equation}
	\tan \phi_1=\frac{d_{y}^\mathrm{R}+d_{x}^\mathrm{I}}{d_{x}^\mathrm{R}-d_{y}^\mathrm{I}},\quad \tan \phi_2=\frac{d_{y}^\mathrm{R}-d_{x}^\mathrm{I}}{d_{x}^\mathrm{R}+d_{y}^\mathrm{I}}.
	\end{equation}
	\par As $\phi_\mathrm{I}$ is a real, continuous and a periodic function of the wave vector $k$, one has,
	\begin{equation}
	\oint_C\partial_k\phi_\mathrm{I}\ dk=0
	\label{eq:zero}
	\end{equation}
	The definition of the winding number, $\nu$, that is the topological invariant that counts the winding of the EPs, can be represented by,\cite{Ghatak_2019}
	\begin{equation}
	\nu_n=\frac{1}{2\pi}\oint_C\partial_k\phi_n\ dk
	\label{eq:w}
	\end{equation}
	where $n=\pm$ is the band index, $\phi_n(k)$ is the argument of the $\mathbf{d}$-vector, that is $\phi_n(k)=\tan^{-1}(d_y/d_x)$ and the contour $C$ denotes the Brillouin zone (BZ), that is, $k$ goes from $-\pi$ to $\pi$.
	It is evident that $\nu$ will only be function of $\phi_\mathrm{R}$ (see Eq.\eqref{eq:zero}).
	From Eqs.\eqref{eq:sum} and \eqref{eq:phi}, $\nu$ splits into two parts, namely,
	\begin{equation}
	\nu=\frac{1}{2}(\nu_1+\nu_2)
	\end{equation}
	where,
	\begin{equation}
	\nu_1=\frac{1}{2\pi}\oint_C\partial_k\phi_1\ dk \quad\mathrm{and}\quad \nu_2=\frac{1}{2\pi}\oint_C\partial_k\phi_2\ dk.
	\end{equation}
	Here, $\phi_1$ is the angle that the line connecting $(d_{x}^\mathrm{R}, d_{y}^\mathrm{R})$ and $(d_{y}^\mathrm{I}, -d_{x}^\mathrm{I})$ make with a line parallel to x-axis (see Fig.\ref{Fig:Locus-EP}), and $\phi_2$ is the corresponding angle for the EP located at $(-d_{y}^\mathrm{I}, d_{x}^\mathrm{I})$.
	So, as shown in Fig.\ref{Fig:Locus-EP}, as $k$ is taken over the BZ, both the EPs will travel along a circle (anti-clockwise), each with radius $\delta_2$ (both denoted by blue circles in Fig.\ref{Fig:Locus-EP}), and their centers are located at $(-\delta_1,0)$ corresponding to the EP at $(d_{y}^\mathrm{I}, -d_{x}^\mathrm{I})$, and at $(\delta_1,0)$ for the EP at $(-d_{y}^\mathrm{I}, d_{x}^\mathrm{I})$.
	The point ($d_{x}^\mathrm{R}$, $d_{y}^\mathrm{R}$) will traverse along a circle (shown by red color in Fig.\ref{Fig:Locus-EP}) of radius $t_2$ whose center is located at ($t_1$, 0).
	This is similar to the locus of the $\pmb{d}$-vector in the $d_x$-$d_y$ plane for the Hermitian version of the model, with the only difference that here we have to deal with the real parts of $d_x$ and $d_y$.
	\begin{figure}[t]
		\includegraphics[width=0.45\textwidth]{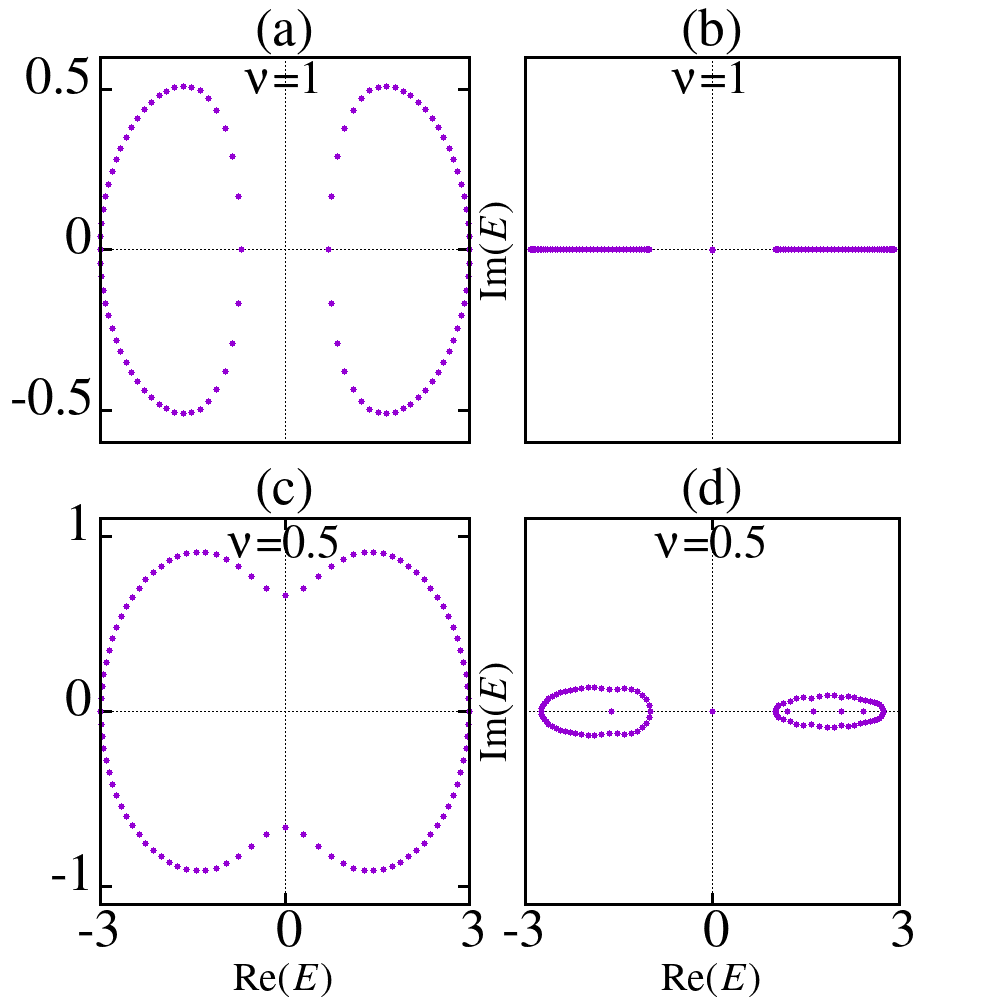}
		\caption{(Color online) Re($E$) vs Im($E$) for (a) $\nu=1$ with PBC, (b) $\nu=1$ with OBC, (c) $\nu=0.5$ with PBC, (d) $\nu=0.5$ with OBC, keeping $t_1=1$ and $t_2=2$. There is a zero energy mode ($|E|=0$) in the right panels.}
		\label{Fig:rm2}
	\end{figure}
	\begin{figure}[h!]
		\includegraphics[width=0.45\textwidth]{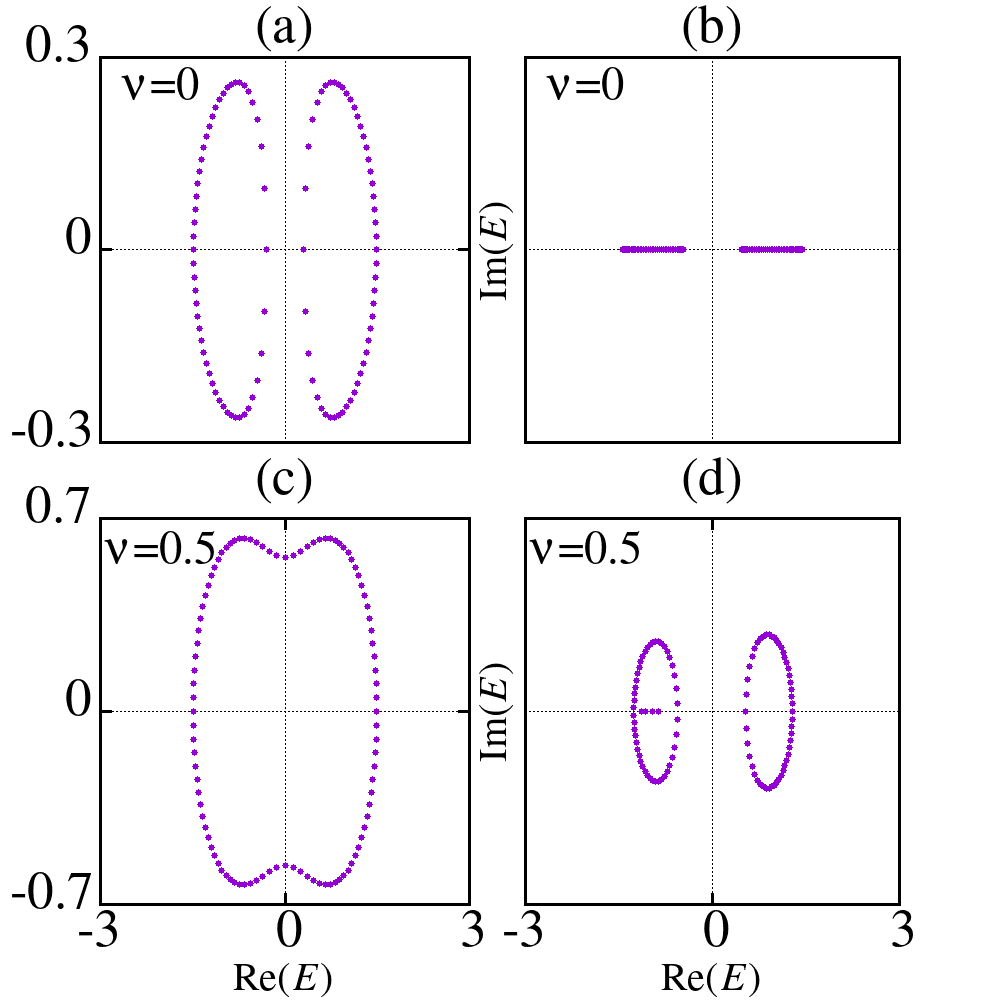}
		\caption{(Color online) Re($E$) vs Im($E$) for (a) $\nu=0$ with PBC, (b) $\nu=0$ with OBC, (c) $\nu=0.5$ with PBC, (d) $\nu=0.5$ with OBC, keeping $t_1=1$ and $t_2=0.5$. The $|E|=0$ mode is missing.}
		\label{Fig:rm1}
	\end{figure}
	\par We can now construct the phase diagrams in Fig.\ref{Fig:PhaseDiagrams-nu} for the winding number $\nu$ in the plane defined by the non-reciprocity parameters, $\delta_1$ and $\delta_2$ for both $t_{1} > t_{2}$ and $t_{1} < t_{2}$. 
	There are three distinct regions in these phase diagrams, where $\nu$ assumes values $0$, $\frac{1}{2}$ and $1$ corresponding to zero winding, winding one of the set of EPs, and both the EPs respectively, as $k$ is taken from over the BZ.
	Thus, there are clear evidences of phase transitions from one topological phase to another, characterized by the winding number discontinuously changing from $1$ to $0.5$, or from a topological to a trivial phase where the winding number jumps from $0$ to $0.5$.
	These abrupt changes are depicted as a function of $\delta_{2}$ in the lower panel of Fig.\ref{Fig:PhaseDiagrams-nu} along the straight lines (in red) shown in its upper panel.	
	\begin{figure*}[t]
		\includegraphics[width=1\textwidth]{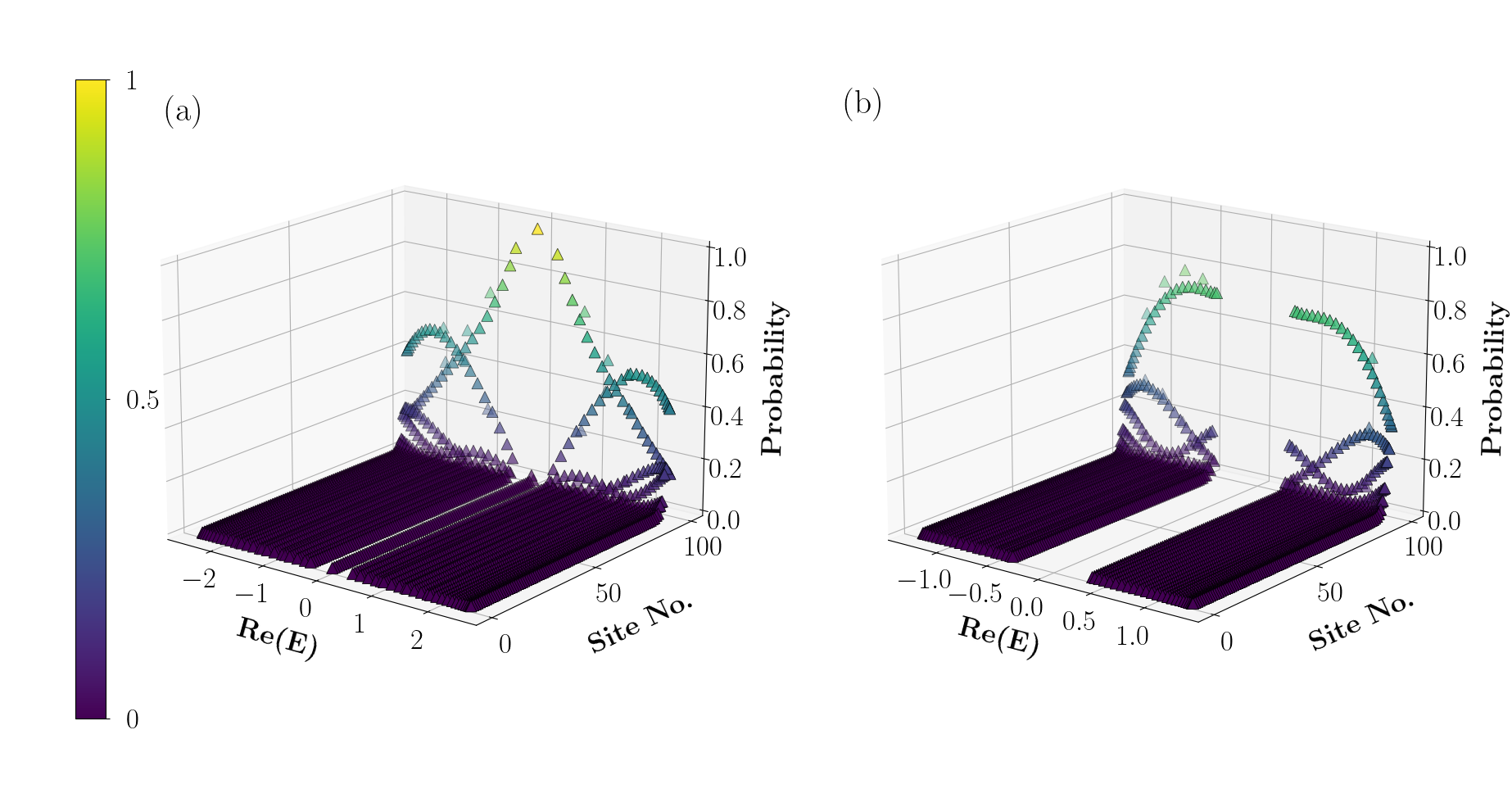}
		\caption{(Color online) NHSE in the  non-$\mathcal{PT}$ symmetric model. $\mathrm{(a)}\;t_1=1$, $t_2=2$, $\delta_1=0.5$, $\delta_2=1.3$ and $\mathrm{(b)}\;t_1=1$, $t_2=0.5$, $\delta_1=0.5$, $\delta_2=0.3$. The system comprises of $100$ lattice sites.}\centering
		\label{fig:1}
	\end{figure*}
	\par To understand the topological phase transitions more succinctly, we plot the band structure in Fig.\ref{Fig:b1}, that is, the real part of the energy, corresponding to the expression $E(k)$ given by,
	\begin{align}
		E_{\pm}=\pm\bigg[t_1^2+t_2^2-\delta_1^2-\delta_2^2+&2(t_1t_2+\delta_1\delta_2)\cos k\nonumber\\-&2i(t_1\delta_2+t_2\delta_1)\sin k\bigg]^{\frac{1}{2}}
		\label{eq:non}
	\end{align} for two representative values of $\delta_2$ that denote two different values of the winding number, $\nu$ corresponding to both $t_{1} > t_{2}$ and $t_{1} < t_{2}$ in the upper and the lower panels of respectively.
	For $t_{1} < t_{2}$, $\delta_1 = 0.1$ corresponds to $\nu = 1$, where we have observed the spectral gap being non-zero everywhere in BZ.
	Thus it makes sense to conclude that the real part energy shows a topological gap with $\nu=1$ for $t_{1} < t_{2}$, and a trivial gap with $\nu=0$ for $t_{1} > t_{2}$, while $\nu=0.5$ corresponds to gap closing scenario for both the cases.
	However, there is an important difference between the two.
	The zero mode for OBC which exists for $t_1<t_2$, is absent for $t_1>t_2$.
	\begin{figure}[h]
		\centering
		\includegraphics[width=0.45\textwidth]{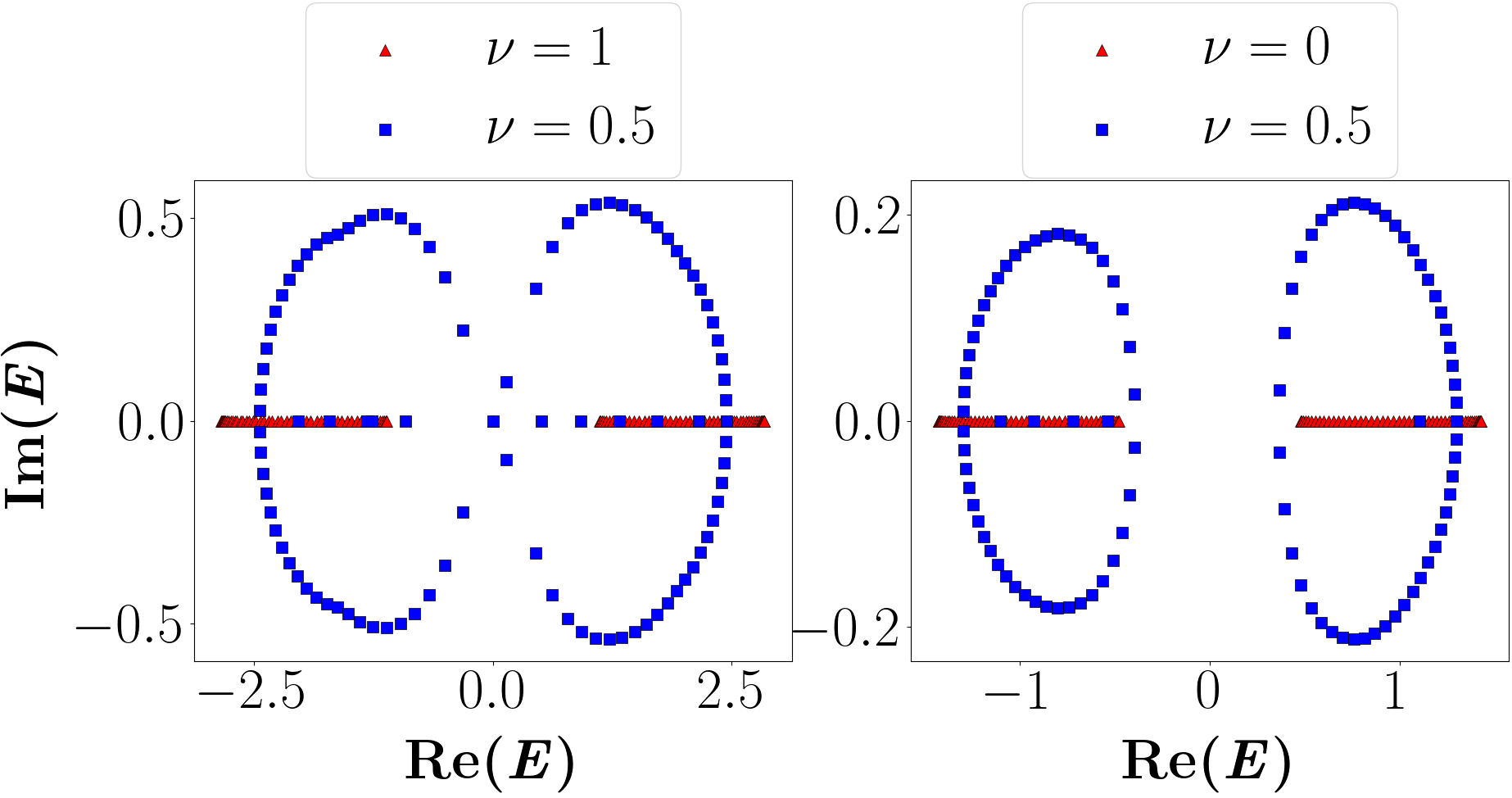}
		\caption{(Color online) Re($E$) vs Im($E$) graphs for non-$\mathcal{PT}$ symmetric  model with OBC with $\mathrm{(a)}\;t_1=1$, $t_2=2$, $\delta_1=0.5$, $\delta_2=0.3$ for $\nu=1$ and $\delta_1=0.6$, $\delta_2=1.3$ for $\nu=0.5$, and $\mathrm{(b)}\;t_1=1$, $t_2=0.5$, $\delta_1=0.3$, $\delta_2=0.15$ for $\nu=0$ and $\delta_1=0.5$, $\delta_2=0.3$ for $\nu=0.5$.}
		\label{fig:3}
	\end{figure}
	\begin{figure*}[t]
		\includegraphics[width=0.45\textwidth]{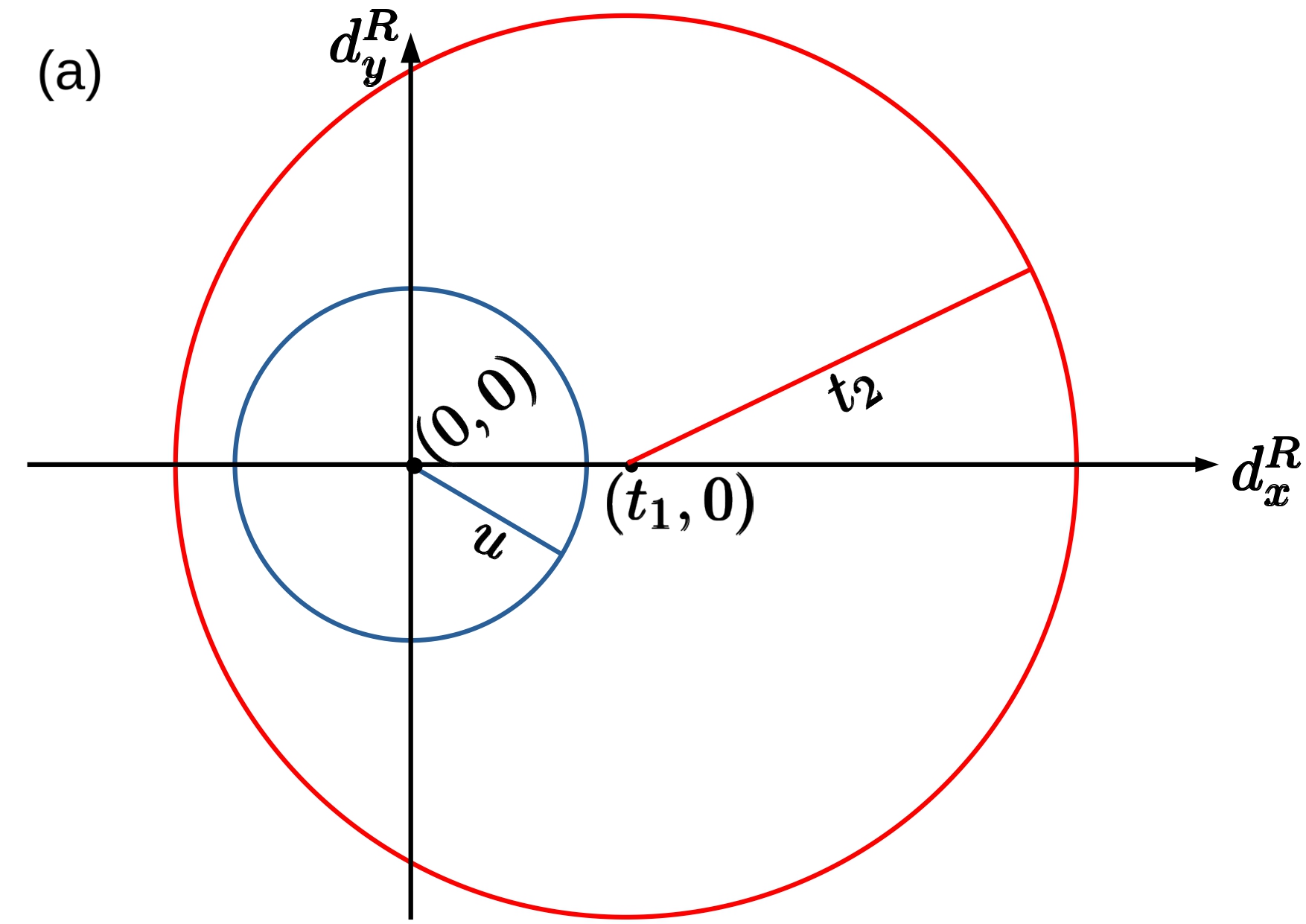}\hfill
		\includegraphics[width=0.45\textwidth]{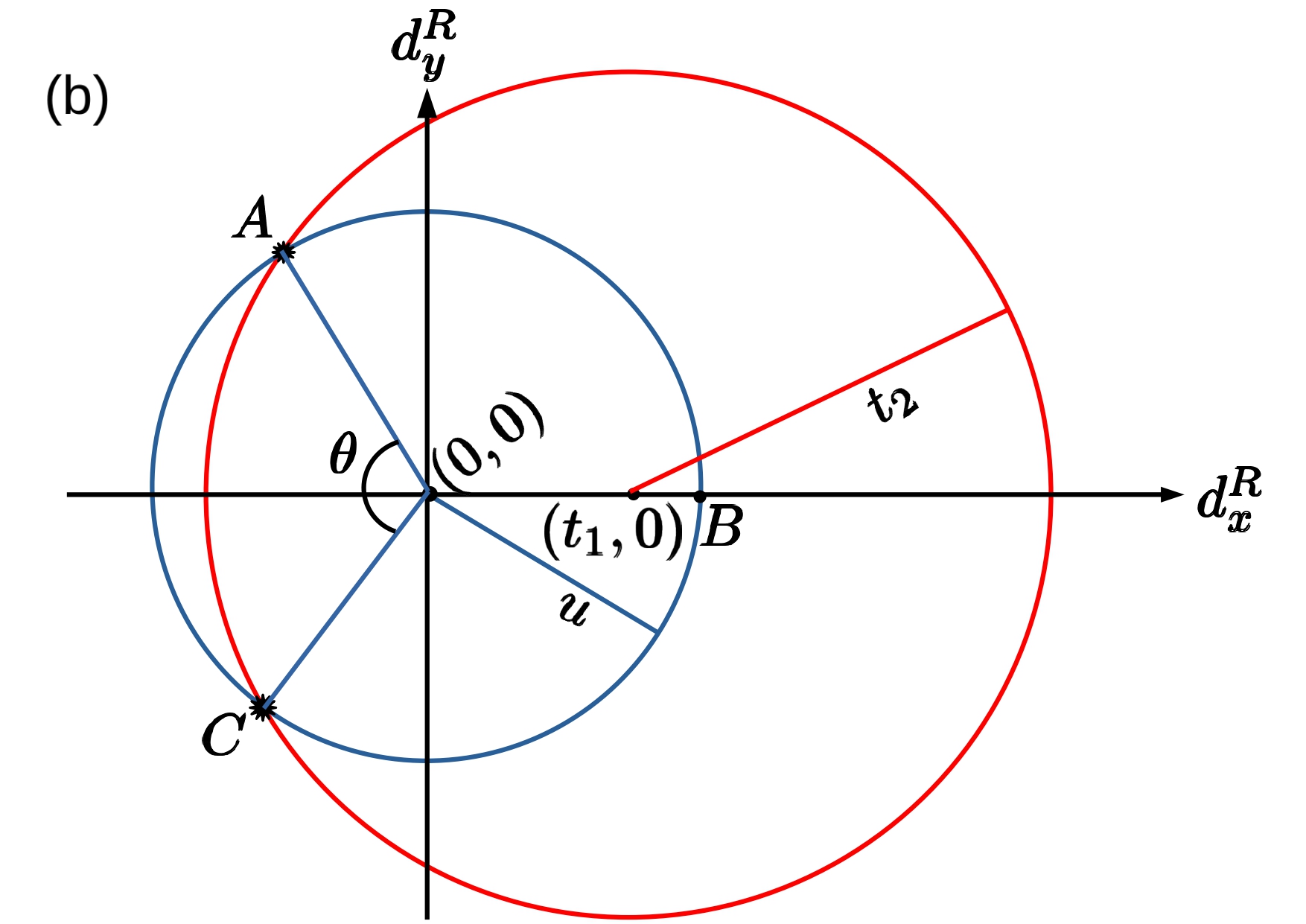}\vskip 0.1 in
		\includegraphics[width=0.45\textwidth]{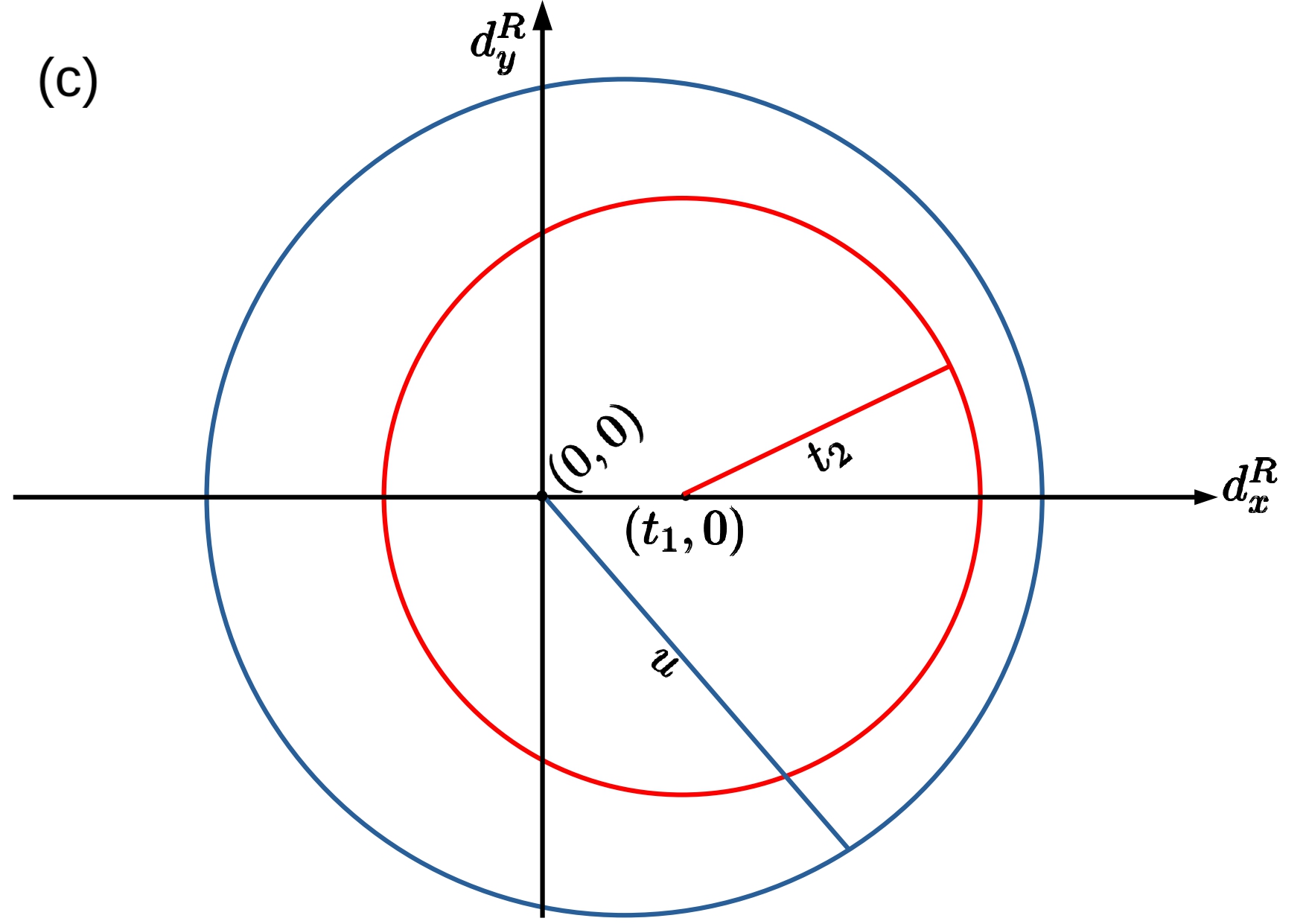}\hskip 1.3 in\includegraphics[width=0.3\textwidth]{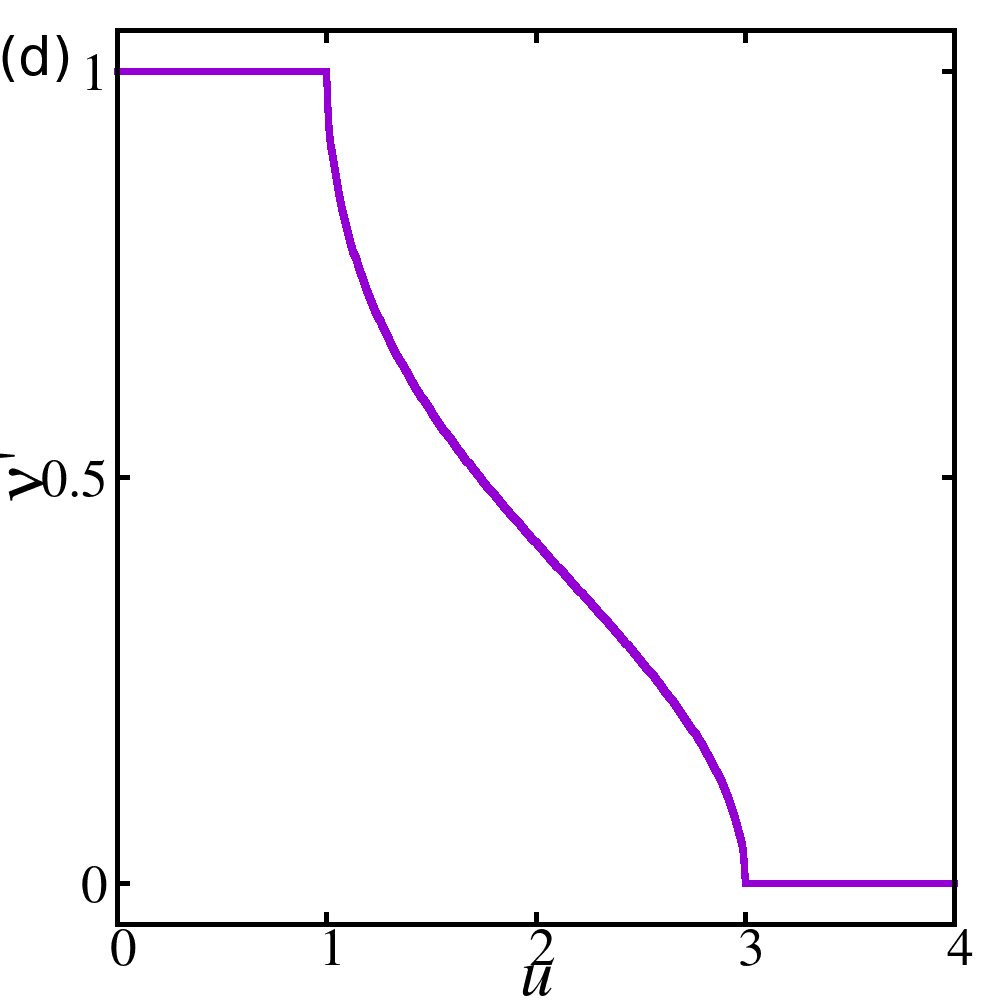}~~~~~~~~~~~
		\caption{(Color online) (a) The EC is completely surrounded by the red circle, (b) the EC is partially surrounded by the red circle, (c) the red circle is fully surrounded by the EC, (d) winding number vs potential strength, $u$, for $t_1=1,\ t_2=2$.}
		\label{fig:EC1}
	\end{figure*}
	Further, the discontinuous transitions of one value of $\nu$ to another occur when Eq.\eqref{eq:ep2} is satisfied.
	For example, in Fig.\ref{Fig:PhaseDiagrams-nu}, $\nu$ drops discontinuously from $1$ to $0.5$ at $\delta_2=0.5$ for $t_1=1$, $t_2=2$, and from $0$ to $0.5$ at $\delta_2=0.25$ for $t_1=1$, $t_2=0.5$.

	\par Next we distinguish between the behavior of the imaginary part of the energy as a function of its real part for both the PBC and the OBC corresponding to $t_{1} < t_{2}$ (Fig.\ref{Fig:rm2}) and $t_{1} > t_{2}$ (Fig.\ref{Fig:rm1}).
	The eigenvalues always come in pairs with `$+$' and `$-$' signs, that is there must be some $-E$ for every $E$ due to Hamiltonian's chiral nature (obeys Eq.\eqref{eq:c}).
	In the upper panel of Fig.\ref{Fig:rm2}, we contrast between the PBC and OBC for $\nu=1$.
	Apart from demonstrating a completely different behavior, where the energy has both real and imaginary parts for the PBC, while Im($E$)$=0$ for OBC, we get a couple of zero energy modes ($|E|=0$).
	At $\nu=0.5$, although Im($E$)$\ne0$ in both the PBC and the OBC, the zero energy modes continue to exist for the OBC.
	The above scenario suffers drastically, that is $|E|$ remains non-zero all the while for $t_{1} > t_{2}$ (Fig.\ref{Fig:rm1}), albeit there are differences noted in the behavior of system in the PBC and the OBC.
	Thus the situation corresponding to $\nu=0.5$, although with a gapless band structure (Re($E$)$=0$) at the edges of the BZ for both $t_1<t_2$ and $t_1>t_2$ (see right panel of Fig.\ref{Fig:b1}), are distinct.
	\par Let us have a closer look at Figs.\ref{Fig:rm2}(d) and \ref{Fig:rm1}(d).
	They seem to form closed loops from a distance, while a careful inspection reveals that these are not really closed loops.
	For the sake of clarity, Figs.\ref{Fig:rm2}(b) and \ref{Fig:rm2}(d) are zoomed in Fig.\ref{fig:3}(a).
	It can be seen in Fig.\ref{fig:3}(a) that though two loops form at both sides of the imaginary axis for $\nu=0.5$, some points scatter on the real axis, suggesting that $\mathrm{Im}(E)=0$, (with a couple of eigenvalues with $|E|$ being zero because of $t_1<t_2$), making it more of a scattered plot, rather than a closed loop.
	Also it must be noted that for $\nu=1$, there are no loops formed.
	The points lie only on the real axis, making all the eigenvalues real.
	\par Similarly Fig.\ref{fig:3}(b) is a clearer picture of Figs.\ref{Fig:rm1}(b) and \ref{Fig:rm1}(d).
	Here also some points scatter on the real axis apart from making loops at each side of the imaginary axis for $\nu=0.5$.
	It must be noted that for $\nu=0$, no loops are formed here as well.
	Unlike the previous case (Fig.\ref{fig:3}(a) with $\nu=0.5$), here there are no eigenvalues with $\mathrm{abs}(E)=0$ for $t_1>t_2$.
	These so called "closed loops" are distinct from the closed loops we obtained in case for PBC (Figs.\ref{Fig:rm2}(c) and \ref{Fig:rm1}(c)).
	\par In fact, a common perception is that closed loops only occur in PBC.
	However, OBC spectrum does form closed loops as well.
	In the work of Liu $\mathit{et\;al.}$\cite{loc} on the non-Hermitian Aubry Andr\'{e} model, it can be seen that system with OBC forms closed loops as shown in Figs.9(a)-(c) of the paper.
	\par Let us now describe the NHSE in the non-$\mathcal{PT}$ symmetric system.
	The NHSE is depicted in Fig.\ref{fig:1}.
	Majority of the eigenstates are localized at either of the edges depending upon the signs of $\delta_1$ and $\delta_2$.
	We have showed the NHSE for $t_1=1$ and two values of $t_2$, namely, $0.5$ and $2$ with $\nu=0.5$ in Fig.\ref{fig:1}.
	It can be seen almost all the eigenstates for both the cases are localized at the edges of the system with the zero energy modes present for the case $t_1<t_2$ (left panel of Fig.\ref{fig:1}) and absent for $t_1>t_2$ (right panel of Fig.\ref{fig:1}).
	If we make any of $\delta_1$ or $\delta_2$ negative, the eigenstates will localize at the other end, namely, the left edge of the system.
	Same phenomena (NHSE) have been observed for $\nu=1$ with $t_1<t_2$ and $\nu=0$ with $t_1>t_2$ also.
	\par We have also studied two different types of the spectral gaps, such as the point gap and the line gap from the energy spectra of the Hamiltonian with PBC.
	The energy gaps play a vital role in determining the topological properties of the system\cite{PhysRevX.9.041015}.
	There are further two kinds of line gaps, namely, line gaps with respect to the real axis ($L_r$), and gap with respect to the imaginary axis ($L_i$).
	In Fig.\ref{Fig:PhaseDiagrams-nu}(a), which defines $t_1<t_2$, the `sky blue' regions correspond to point gap , while the `light magenta' region hosts line gap.
	This means that we have $P$ type energy gap when $\nu=0.5$ (shown in Fig.\ref{Fig:rm2}(c)) and $L_r$ type gap when $\nu=1$ (shown in Fig.\ref{Fig:rm2}(a)).
	Similarly for $t_1>t_2$, shown in Fig.\ref{Fig:PhaseDiagrams-nu}(b), we have $P$ type energy gap for $\nu=0.5$ (shown in Fig.\ref{Fig:rm1}(c)), which is shown by `sky blue' region and $L_r$ type gap when $\nu=0$ (shown in Fig.\ref{Fig:rm1}(a)) shown by `white' region in Fig.\ref{Fig:PhaseDiagrams-nu}(b).
	These results are consistent with the band structure (Re ($E$) vs $k$) graphs (shown in Fig.\ref{Fig:b1}), which show gapped spectrum for $\nu=1$ (for $t_1<t_2$) and $\nu=0$ (for $t_1>t_2$), whereas the spectrum is gapless for $\nu=0.5$ for both $t_1<t_2$ and $t_1>t_2$.
	These transitions occur when there is a jump in the value of the winding number that are denoted by the continuous bold lines in Figs.\ref{Fig:PhaseDiagrams-nu}(a) and \ref{Fig:PhaseDiagrams-nu}(b).
	This feature suggests of a connection between the topology of the system and the complex energy spectrum.
	\subsection{$\mathcal{PT}$ symmetric case}
	We shall explore an alternate route for rendering non-Hermiticity to our dimerized Hamiltonian, which may be achieved via an imaginary potential, $u$, that is given by $h_2(k)$ in Eq.\eqref{eq:u}.
	The energy eigenvalues of $h_2(k)$ are given by (see Eq.\eqref{eq:eig}),
	\begin{equation}
	E_{\pm}=\pm\sqrt{|t_1+t_2e^{-ik}|^2-u^2}.
	\end{equation}
	The right eigenvectors of $h_2(k)$ are given by,
	\begin{equation}
		\left|\Psi_{2+}^{RE}\right>=\begin{pmatrix}
			e^{-i\phi_k}\cos \theta_k\\ \sin\theta_k	\end{pmatrix}\quad\mathrm{and}\quad\left|\Psi_{2-}^{RE}\right>=\begin{pmatrix}
		-e^{-i\phi_k}\sin \theta_k\\ \cos\theta_k
		\end{pmatrix}
		\label{eq:RE_u}
	\end{equation}
	where, $\phi_k=i\ln\left|\frac{t_1+t_2e^{-ik}}{|t_1+t_2e^{-ik}|}\right|$ and $\theta_k=\tan^{-1}\left(\sqrt{\frac{E_+-iu}{E_++iu}}\right)$.
	\begin{figure}[h]
		\includegraphics[width=0.5\textwidth,height=0.25\textwidth]{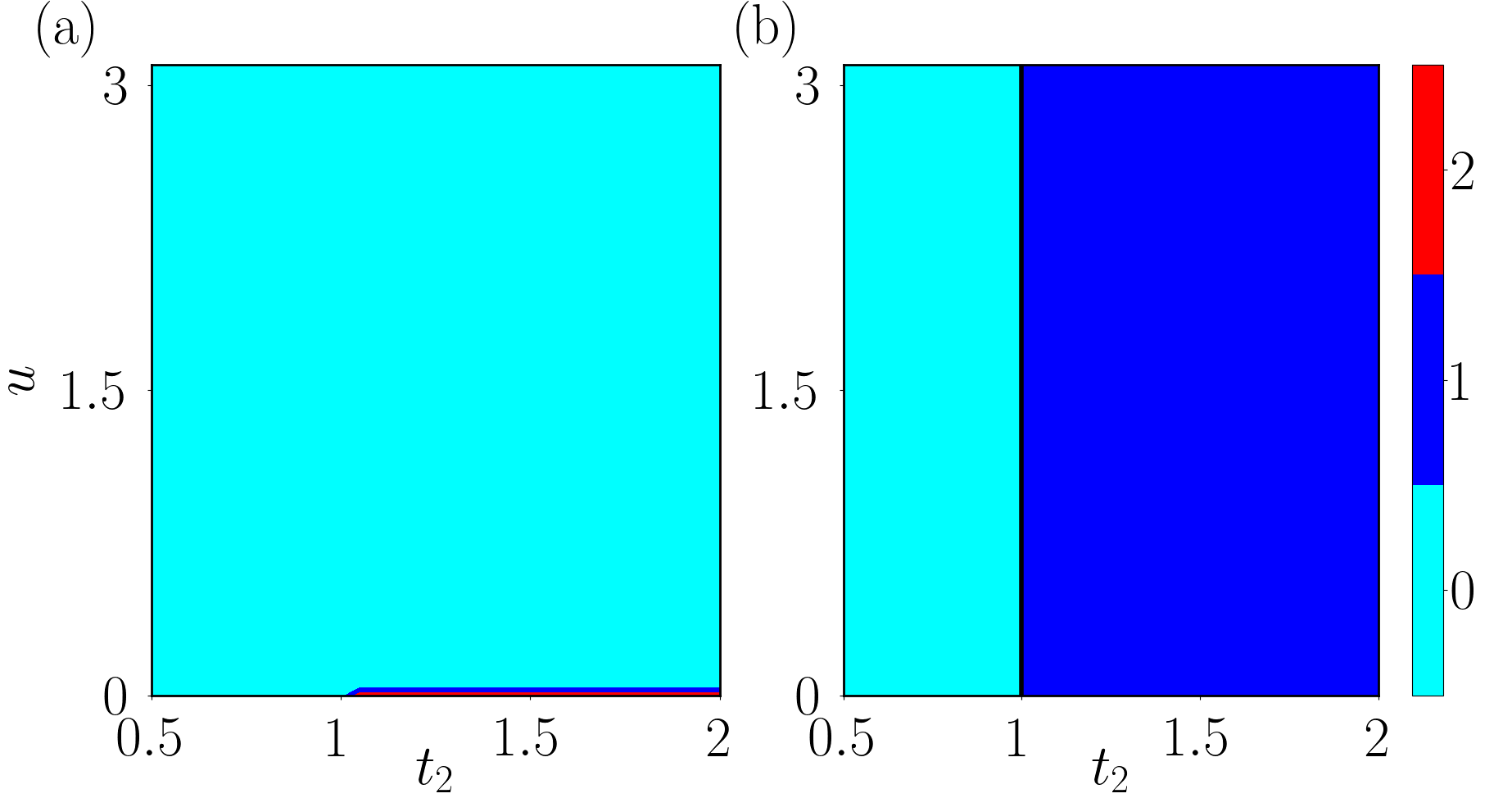}
		\caption{(Color online) (a) Phase diagram of the number of zero energy modes ($|E|=0$) drawn in $u$-$t_2$ parameter space. (b) The phase diagram of winding number ($\nu$) using Eq.\eqref{eq:w} in the $u$-$t_2$ space keeping $t_1$ fixed at $1$.}\label{fig:phase_u}
	\end{figure}
	The topological invariant for the $\mathcal{PT}$ symmetric model is considered as the usual winding number, defined in Eq.\eqref{eq:w}.
	For this case, expressions for $d_x$ and $d_y$ are given by Eq.\eqref{eq:du}.
	Fig.\ref{fig:phase_u}(a) shows that there are no zero energy modes ($|E|=0$) as long as $u\ne0$.
	These modes (a couple of them) are restored as soon as $u$ becomes $0$ and $t_1$ becomes smaller than $t_2$ (shown by thin red part at the bottom of Fig.\ref{fig:phase_u}(a)), which is a known result for the Hermitian SSH model.
	Even though there are no abs$(E)=0$ modes for $t_1<t_2$, we still get $\nu=0$ for $t_1>t_2$, and $\nu=1$ for $t_1<t_2$ irrespective of the values of $u$ (shown in Fig.\ref{fig:phase_u}(b)) because of the absence ($t_1>t_2$) and presence ($t_1<t_2$) of edge modes.
	\begin{figure}\hspace{-0.3 in}[t]
		\includegraphics[width=0.4\textwidth, height=0.64\textwidth]{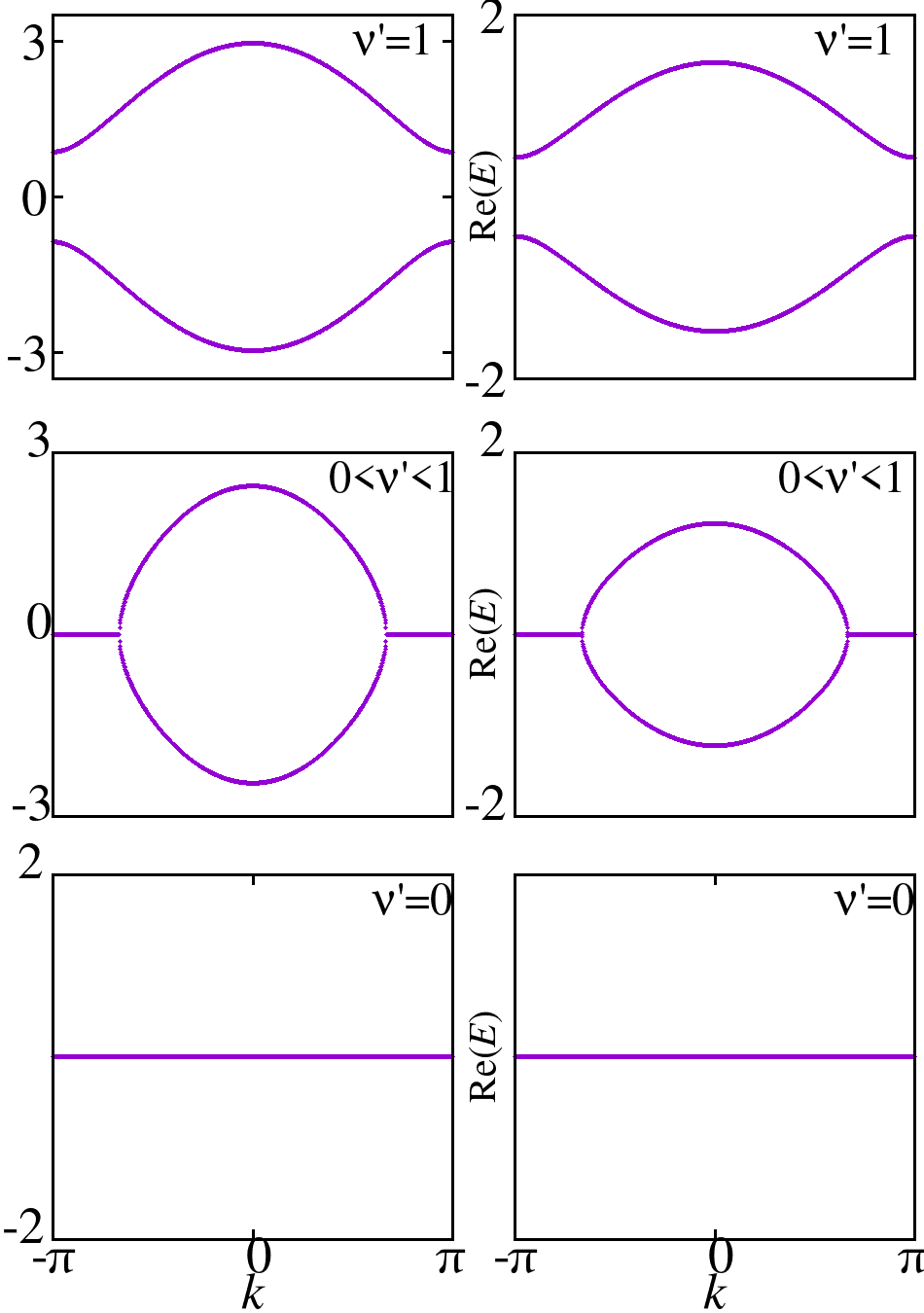}
		\caption{(Color online) The band structure (Re($E$)) for $t_2=2$ (first column) and $t_2=0.5$ (second column) with $t_1=1$ corresponding to different values of the winding number $\nu$.}\label{fig:b2}
	\end{figure}
	\begin{figure}[t]
		\includegraphics[width=0.45\textwidth]{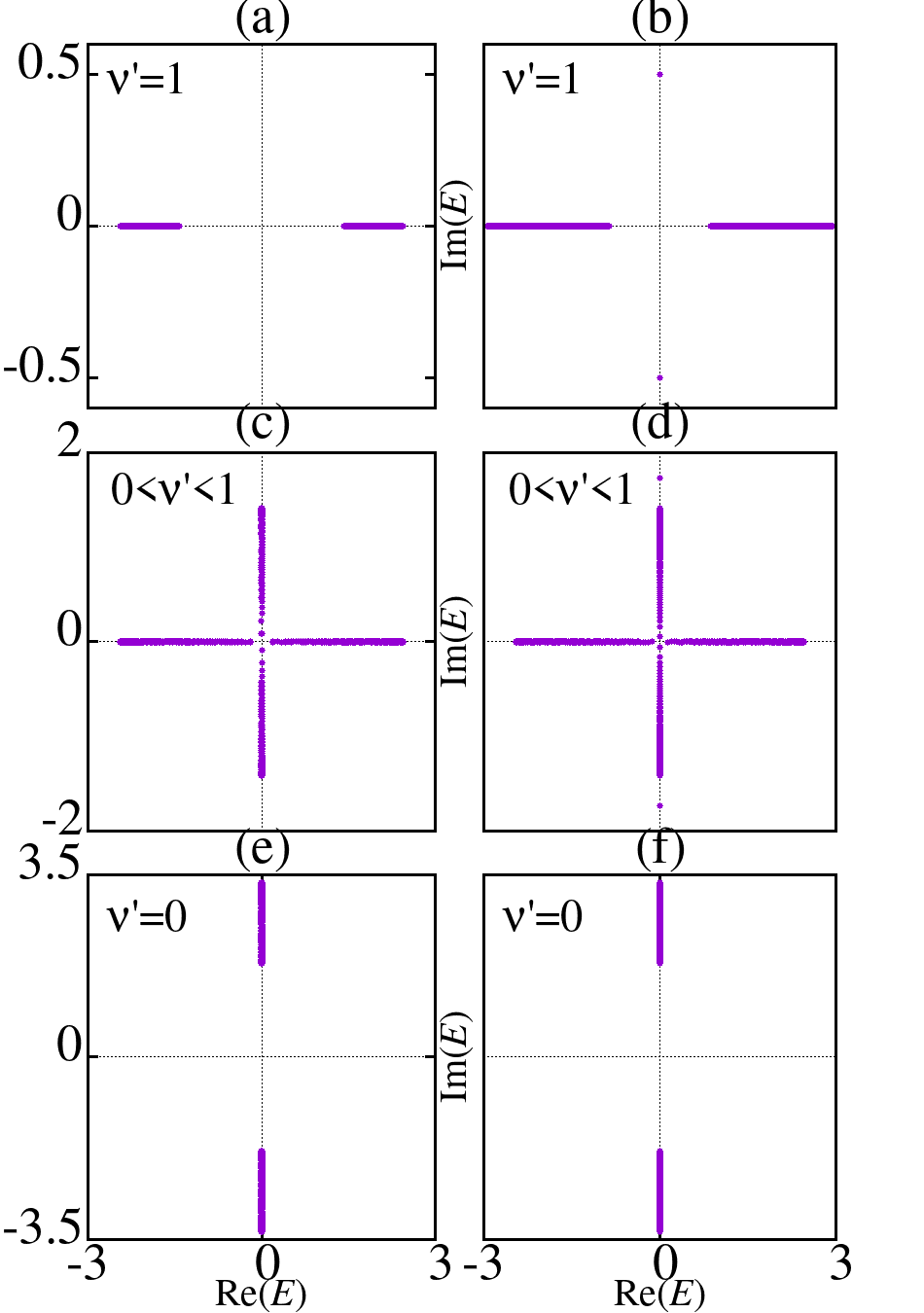}
		\caption{(Color online) Re($E$) vs Im($E$) with PBC (first column) and OBC (second column) for three cases: (a) and (b) correspond to $u=0.5$, (c) and (d) correspond to $u=2$, (e) and (f) correspond to $u=3.5$, keeping $t_1=1$ and $t_2=2$.}\label{fig:rm3}
	\end{figure}
	\begin{figure}[t]
		\includegraphics[width=0.45\textwidth]{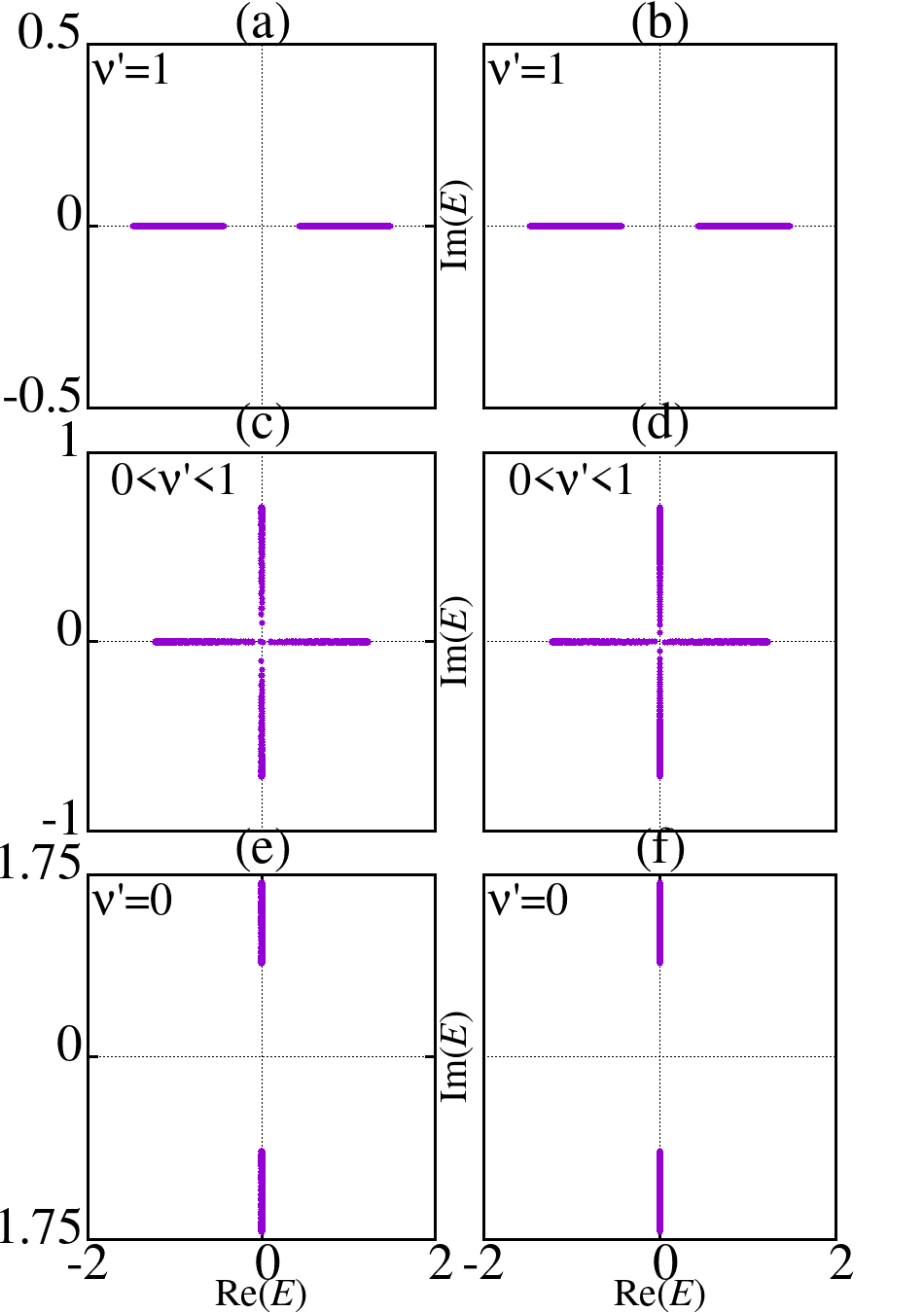}
		\caption{(Color online) Re($E$) vs Im($E$) with PBC (first column) and OBC (second column) for three cases: (a) and (b) correspond to $u=0.25$, (c) and (d) correspond to $u=1$, (e) and (f) correspond to $u=1.75$, keeping $t_1=1$ and $t_2=0.5$.}\label{fig:rm4}
	\end{figure}
	\par We have also computed the complex Berry phase for this model.
	The complex Berry phase is an important quantity in the $\mathcal{PT}$ symmetric case, and is given by,\cite{PhysRevB.97.045106}
	\begin{equation}
		Q^c_n=i\int_{BZ}\Bigg<\lambda^n_k\Bigg|\frac{\partial}{\partial k}\Bigg|\psi^n_k\Bigg>d\\k
	\end{equation}
	where $\bra{\lambda^n_k}$ and $\ket{\psi^n_k}$ are the left and the right eigenvectors respectively corresponding to $n^{\mathrm{th}}$ band respectively.
	To make our presentation complete, we have included a complete derivation of the complex Berry phase in Appendix A.
	The complex Berry phase exhibits the same behavior as that shown by the winding number ($\nu$), where the former is either $2\pi$ and $0$ for $t_1<t_2$ and $t_1>t_2$ respectively, while $\nu$ takes a value $1$ as long as $t_1<t_2$, and becomes $0$ when $t_1>t_2$,  irrespective of $u$.
	In this way, the complex Berry phase behaves similar to that of $\nu$, although the value has to be multiplied by $2\pi$.
	\par Now let us focus on finding the EPs of this system.
	The energy eigenvalues coalesce when $E_{\pm}$ become zero, which leads to,
	\begin{equation}
	|t_1+t_2e^{-ik}|^2=u^2
	\label{eq:exp}
	\end{equation}
	The coalescence of the eigenvectors demands also Eq.\eqref{eq:exp} to be satisfied, and thus is in contrast with the non-$\mathcal{PT}$ symmetric case, where the conditions corresponding to the coalescence of the eigenvalues and the eigenvectors are distinct.
	In terms of components of the $\pmb{d}$-vector, Eq.\eqref{eq:exp} can be written as,
	\begin{gather}
	(d_{x}^\mathrm{R})^2+(d_{y}^\mathrm{R})^2=u^2
	\label{dz}
	\end{gather}
	which is the equation of a circle with radius $u$ in a space spanned by $d_{x}^\mathrm{R}-d_{y}^\mathrm{R}$.
	Expressions for $d_{x}^\mathrm{R}$ and  $d_{y}^\mathrm{R}$ can be obtained from Eq.\eqref{eq:du}.
	So the EPs are arranged on the circumference of a circle of radius $u$, and there are infinite number of EPs which reside on the circumference.
	A schematic diagram is shown, via blue circle, while the red circle (of radius $t_2$) denotes the locus of the $\pmb{d}$-vector in the $d_{x}^\mathrm{R}(k)-d_{y}^\mathrm{R}(k)$ plane, in Fig.\ref{fig:EC1}.
	\begin{figure*}[t]
		\includegraphics[width=\textwidth]{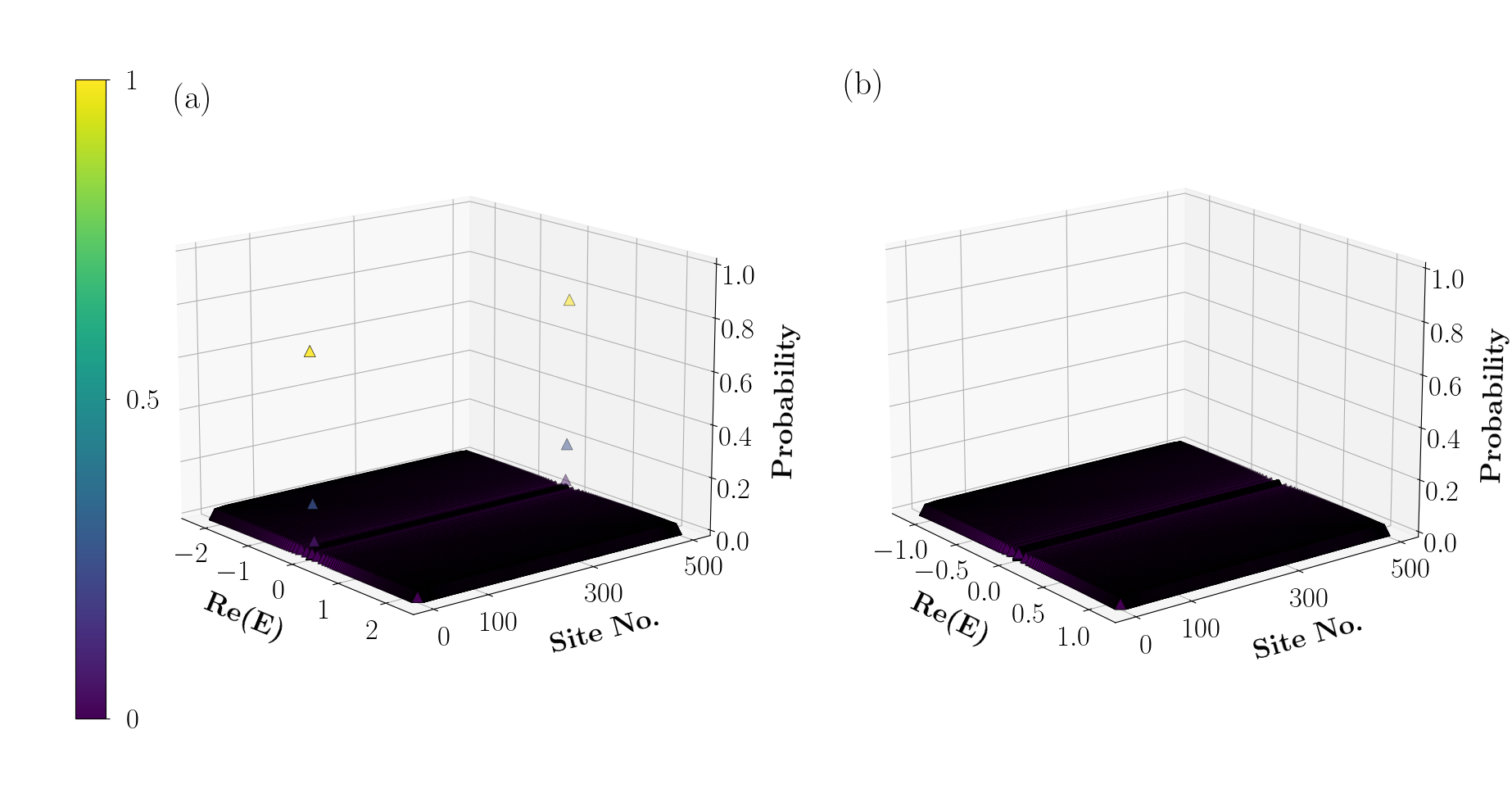}
		\caption{(Color online) No NHSE in the $\mathcal{PT}$ symmetric model. $\mathrm{(a)}\;t_1=1$, $t_2=2$, $u=2$ and $\mathrm{(b)}\;t_1=1$, $t_2=0.5$, $u=1$.}\centering 
		\label{fig:2}
	\end{figure*}
	\par Now, we can define a new winding number ($\nu'$) for this system in the following way.
	$\nu'$ is the ratio of the portion of the $\mathit{Exceptional\ Circle}$ in blue color (will be referred as EC hereafter) wound by the $d_{x}^\mathrm{R}(k)-d_{y}^\mathrm{R}(k)$ curve, shown in red color, and the circumference of the EC. 
	We shall show the calculations of $\nu'$ afterwards.
	For example, as shown in Fig.\ref{fig:EC1}(b), $\nu'$ is denoted by the ratio of the arc $ABC$ to the whole EC.
	Thus $\nu'$ may assume any value between $0$ and $1$ depending on the portion of the overlap of EC with the red circle.
	This is true for any value of the potential $u$ (refer to Fig.\ref{fig:EC1}).
	\par It may be noted that we have kept $t_1$ fixed and considered two values of $t_2$, namely $2$ and $0.5$, which fix the radius of the red circle at these values.
	Let us first discuss $t_1<t_2$.
	In order to compute the overlap of the EC with the red circle, there are two possibilities.
	Either the EC is completely within the red circle (Fig.\ref{fig:EC1}(a)), which corresponds to $\nu'=1$, or to enumerate the partial overlap we compute the angle $\theta$ in Fig.\ref{fig:EC1}(b) (shown below).
	This corresponds to $\nu'<1$ (in this case $\nu'$ is close to $1$).
	In Fig.\ref{fig:EC1}(c), the red circle is completely inside EC which corresponds to zero overlap, and hence $\nu'=0$.
	We show the variation of $\nu'$ as a function of $u$ in Fig.\ref{fig:EC1}(d).
	Till $u=1$ ($=|t_1-t_2|$), $\nu'$ remains at a value $1$.
	For $1<u<3$, we get $0<\nu'<1$, while for $u>3$ ($=t_1+t_2$), $\nu'$ remains at a value $0$.
	The above scenario can be contrasted with the case $t_1>t_2$, except for the red circle and the EC are apart with zero overlap for the first case of $t_1<t_2$ (Fig.\ref{fig:EC1}(a)), that is, when $u<t_2-t_1$.
	$\nu'$ takes the value $1$ there and behaves similarly of that of the previous case ($t_1<t_2$) afterwards.
	\par Now let us return back to the calculations that yield the plots shown in Fig.\ref{fig:EC1}(d).
	At the points of overlap of these two circles (EC and the red circle), if we substitute,
	\begin{equation*}
	d_{x}^\mathrm{R}=t_1+t_2\cos k, \quad d_{y}^\mathrm{R}=t_2\sin{k}
	\end{equation*}
	in Eq.\eqref{dz}, we shall obtain,
	\begin{gather}
	u^2-2d_{x}^\mathrm{R}t_1+t_1^2=t_2^2,\quad\mathrm{and}\quad d_{x}^\mathrm{R}=\frac{u^2+t_1^2-t_2^2}{2t_1}
	\end{gather}
	Thus substituting $d_{x}^\mathrm{R}$ in Eq.\eqref{dz}, one gets,
	\begin{equation}
	d_{y}^\mathrm{R}=\pm\frac{1}{2t_1}\sqrt{\left((u+t_1)^2-t_2^2\right)\left(t_2^2-(u-t_1)^2\right)}
	\end{equation}
	$d_{y}^\mathrm{R}$ must be real, and for that to happen, we should have,
	\begin{equation}
	|t_1-t_2|\leq u\leq t_1+t_2
	\label{eq:real}
	\end{equation}
	It can be shown that the angle subtended by the arc $ABC$ at the center of EC, shown by $(0,0)$ in Fig.\ref{fig:EC1}(b), is given by,
	\begin{equation}
	\theta=2u\tan^{-1}\left[\frac{\sqrt{\Big((u+t_1)^2-t_2^2\Big)\Big(t_2^2-(u-t_1)^2\Big)}}{u^2+t_1^2-t_2^2}\right]
	\end{equation}
	The value of $\theta$ is such that, $\theta\in\big[-\pi,\pi\big]$.
	Thus the winding number, which is denoted by the overlap of EC with the red circle is given by, 
	\begin{gather}
	\nu'=\frac{\mathrm{arc length}(ABC)}{2\pi u}
	\end{gather}
	This finally yields the expression for the winding number as,
	\begin{equation}
	\nu'=\frac{1}{\pi}\tan^{-1}\left[\frac{\sqrt{\Big((u+t_1)^2-t_2^2\Big)\Big(t_2^2-(u-t_1)^2\Big)}}{u^2+t_1^2-t_2^2}\right]\label{eq:w'}
	\end{equation}
	It is clear from Eq.\eqref{eq:w'}, that this winding number, $\nu'$, can take all possible values between $0$ and $1$.
	\par The band structure (Re($E$) vs $k$) is plotted in Fig.\ref{fig:b2}.
	The band gap closes at the point, $u=|t_1-t_2|$, which happens when the EC touches the $d_{x}^\mathrm{R}-d_{y}^\mathrm{R}$ curve internally, and remains so until a value $u=t_1+t_2$, where Re($E(k)$) becomes zero.
	Similar phenomena happen for both $t_1<t_2$ and $t_1>t_2$.
	In order to distinguish the band structures for the two dimerized cases, $t_1<t_2$ (left panel) and $t_1>t_2$ (right panel), we have plotted Re($E$) vs $k$ in Fig.\ref{fig:b2} corresponding to $\nu'=0$ and $\nu'\ne 0$ respectively.
	The top row shows results for small $u$ where $\nu'=1$ for both $t_1<t_2$ and $t_1>t_2$.
	The spectrum is gapped in each case, however the nature of the gaps are different, since they correspond to different values of $\nu'$.
	The central panel of Fig.\ref{fig:b2} denotes the band structure for intermediate values of $u$ where we have $0<\nu'<1$ in each case.
	The situation corresponds to the closure of the spectral gap occurring at $k$-values away from the edge of the BZ, along with the presence of zero modes (Re($E$)$=0$) in both the cases.
	The presence of the spectral gap in certain parts of the BZ probably bears the testimony of a finite winding number.
	Finally, for large $u$, where $\nu'=0$ in both $t_1<t_2$ and $t_1>t_2$ cases, there are two fold degenerate flat bands at Re($E$)$=0$.
	\par We also plot Im($E$) vs Re($E$) for both the PBC and the OBC corresponding to $t_1<t_2$ and $t_1>t_2$ (Figs.\ref{fig:rm3} and \ref{fig:rm4}).
	The energy eigenvalues always occur in complex conjugate pairs, that is there must be a $E^*$ for every $E$, as the Hamiltonian is $\mathcal{PT}$ symmetric (obeys Eq.\eqref{eq:pt}).
	Some of these eigenvalues are purely real, and the rest are purely imaginary.
	This happens for both the PBC and the OBC.
	However, there is one difference.
	For $t_1<t_2$ with OBC, two edge modes exist with $E=\pm iu$ for all values of the potential $u$, a scenario that is absent in PBC.
	Same phenomena have been observed for $t_1>t_2$, except that there are no edge modes corresponding to either of the energy eigenvalues $E=\pm iu$.\cite{PhysRevB.97.045106}
	The two edge modes continue to exist as long as $t_1<t_2$ with the winding number, $\nu$ being 1 (Fig.\ref{fig:2}), regardless of the value of the staggered imaginary potential strength, $u$.\cite{Ghatak_2019}.
	\par
\begin{figure}[h]
		\includegraphics[width=0.52\textwidth]{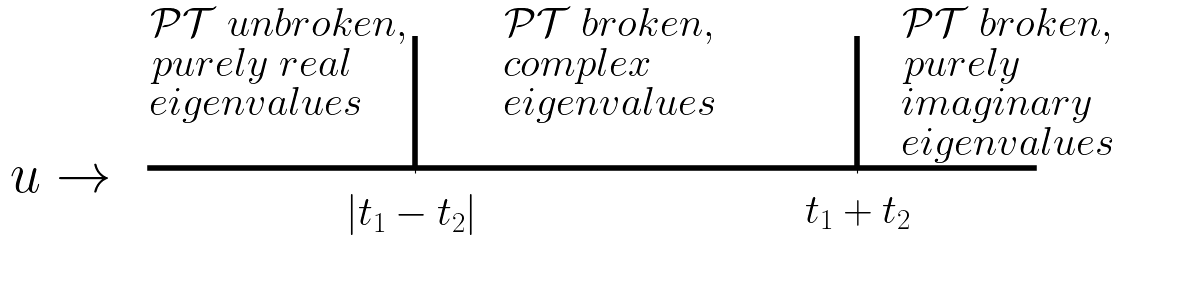}
		\caption{(Color online) The $\mathcal{PT}$ broken and the unbroken regions in the $\mathcal{PT}$ symmetric model are shown as a function of $u$. The boundary between the $\mathcal{PT}$ unbroken region with purely real eigenvalues and $\mathcal{PT}$ broken region with complex eigenvalues is given by $u=|t_1-t_2|$. Also, the boundary separating the $\mathcal{PT}$ broken regions with complex and purely imaginary eigenvalues is given by $u=t_1+t_2$.}
		\label{fig:pt}
	\end{figure}

 Lastly, we discuss the broken and unbroken regions in this model.
	The energy eigenvalues remain purely real till a value of $u$ given by, $u=|t_1-t_2|$.
	Beyond that, the number of purely imaginary eigenvalues increases with $u$, until $u=t_1+t_2$.
	Finally, all the energies become purely imaginary for larger values of $u$.
	So, $u=|t_1-t_2|$ can be thought as the transition point between the $\mathcal{PT}$ broken and the unbroken regions, prior to which the $\mathcal{PT}$ symmetry is unbroken, while after $u=|t_1-t_2|$ it is broken (Fig.\ref{fig:pt}).
	The new winding number, $\nu'$ can be thought of as a measurement of the ratio of the number of energy eigenvalues with real parts being non-zero to the total number of eigenvalues.
	Say, the real parts of $n$ out of $N$ of the energy eigenvalues are non-zero, then $\nu'$ is given by $n/N$.
	\section{Conclusion}
	Here, in this paper, we have investigated the NH SSH model, and examined two separate cases, namely the $\mathcal{PT}$ symmetric case (modeled by a complex on-site potential), and a non-$\mathcal{PT}$ symmetric case (modeled by breaking the reciprocity in the hopping amplitudes).
	The phase diagram of the non-$\mathcal{PT}$ symmetric system encodes a more familiar view of the winding number, which takes integer and half integer values for specific set of the parameters, thereby supporting a phase transition from one value of the winding number to another.
	Different types of spectral gaps are discussed via the real and the imaginary parts of the energies.
	The band structure, given by, Re($E$) vs $k$, supports these transitions via a gap closing scenario.
	A geometrical perspective of understanding the EPs in a more generalized sense has also been provided.
	The EPs for the non-$\mathcal{PT}$ symmetric case are functions of the momentum ($k$) and is thus unusual compared to the Hermitian SSH model.
	For $\mathcal{PT}$ symmetric case, a phase diagram of winding number is obtained and complex Berry phase is being calculated.
	In contrast to the non-$\mathcal{PT}$ symmetric case, where the winding number assumes three distinct values, namely $0$, $\frac{1}{2}$ and $1$, here it can only takes two values, namely $0$ and $1$.
	Another feature emerges as we have infinite number of EPs, arranged on the circumference of a circle, which renders a continuous variation of a newly defined winding number as a function of the strength of the imaginary potential.
	The Re($E$) vs Im($E$) plots show that the eigenvalues are either purely real or purely imaginary (which was not the case for non-$\mathcal{PT}$ symmetric system), which supports the information obtained from the winding number ($\nu'$) and also are in accordance with the band structure Re($E$) vs $k$ plots.
	We also find that the broken and unbroken regions in this case characterized by eigenvalues being real and complex respectively.
	Finally, NHSE is examined for both the cases, which demonstrates significant differences between the non-$\mathcal{PT}$ symmetric and the $\mathcal{PT}$ symmetric cases.
	The variations of the real part of the energy with the corresponding imaginary part are quite distinct in these cases, with only non-$\mathcal{PT}$ symmetric case demonstrating breakdown of BBC, and the occurrence of NHSE. In this case, the occurrence of skin effect depends on the signs of $\delta_1$ and $\delta_{2}$, while the presence or the absence of the zero energy modes depends on whether $t_1<t_2$ or $t_1>t_2$. The $\mathcal{PT}$ symmetric case does not show NHSE, instead only two zero energy edge modes are observed for the topological case ($t_1<t_2$), and none corresponding to the trivial case ($t_1>t_2$) as both the intra-cell and inter-cell hoppings in this model is reciprocal.
	The scenario is similar to the Hermitian SSH model, and hence the conventional BBC is preserved.
	It is prudent to mention that a parallel formalism that uses non-Bloch band theory via complex momenta defined in generalized BZ\cite{PhysRevLett.121.026808} admits usage of the `normal' Bloch wavefunction with real momenta for this scenario.
	
	\begin{appendices}
		\section{: Complex Berry phase}
		The Hamiltonian for the $\mathcal{PT}$ symmetric model in the Bloch form can be written as,
		\begin{gather*}
			H(k)=\begin{pmatrix}
				iu & t_1+t_2e^{-ik}\\
				t_1+t_2e^{ik} & -iu
			\end{pmatrix}.
		\end{gather*}
		The eigenvalues of this Hamiltonian are given by,
		\begin{equation*}
			E_{\pm}=\pm\sqrt{|t_1+t_2e^{-ik}|^2-u^2}.
		\end{equation*}
		In an NH system, the left eigenvector and the right eigenvector satisfy the bi-orthonormal condition given by,
		\begin{equation*}
			\left<\lambda_n|\Psi_m\right>=\delta_{nm}
		\end{equation*}
		where $\left<\lambda_n\right|$ and $\left|\psi_m\right>$ are the left and right eigenvectors corresponding to the eigenvalues $E_n^*$ and $E_m$ respectively.
		Thus, The normalized right eigenvectors are given by,
		\begin{equation}
			\ket{\psi_+}=\begin{pmatrix}
				e^{-i\phi_k}\cos \theta_k\\ \sin\theta_k
			\end{pmatrix}\quad\mathrm{and}\quad\ket{\psi_-}=\begin{pmatrix}
				-e^{-i\phi_k}\sin \theta_k\\ \cos\theta_k
			\end{pmatrix}
			\label{eq:2}
		\end{equation}
		and the left eigenvectors are given by,
		\begin{equation}
			\bra{\lambda_+}=\begin{pmatrix}
				e^{i\phi_k}\cos \theta_k\\ \sin\theta_k
			\end{pmatrix}^T\quad\mathrm{and}\quad\bra{\lambda_-}=\begin{pmatrix}
				-e^{i\phi_k}\sin \theta_k\\ \cos\theta_k
			\end{pmatrix}^T
			\label{eq:3}
		\end{equation}
		where, $\phi_k=i\ln\left|\frac{t_1+t_2e^{-ik}}{|t_1+t_2e^{-ik}|}\right|$ and $\theta_k=\tan^{-1}\left(\sqrt{\frac{E_+-iu}{E_++iu}}\right)$.
		Now the complex Berry phase corresponding to the eigenvalues $E_{\pm}$ is given by,\cite{Ghatak_2019}
		\begin{equation*}
			Q^c_{\pm}=i\int_{BZ}\bra{\lambda^{\pm}_k}\frac{\partial}{\partial k}\ket{\psi^{\pm}_k}d\\k
		\end{equation*}
		Now the global Berry phase is defined via,\cite{PhysRevB.97.045106}
		\begin{equation*}
			Q^c_G=Q^c_++Q^c_-.
		\end{equation*}
		Let us calculate $Q^c_+$ first.
		From Eqs.\eqref{eq:2} and \eqref{eq:3}, we can write $Q^c_+$ as,
		\begin{align*}
			Q^c_+&=i\int_{BZ}\begin{pmatrix}
				e^{i\phi_k}\cos \theta_k\\ \sin\theta_k
			\end{pmatrix}^T\frac{\partial}{\partial k}\begin{pmatrix}
				e^{-i\phi_k}\cos \theta_k\\ \sin\theta_k
			\end{pmatrix}dk \\
			&=i\int_{BZ}\begin{pmatrix}
				e^{i\phi_k}\cos \theta_k\\ \sin\theta_k
			\end{pmatrix}^T \begin{pmatrix}
				\big(-ie^{-i\phi_k}\cos\theta_k\frac{\partial\phi_k}{\partial k}+\\e^{-i\phi_k}\frac{iut_1t_2\sin k}{4E_+^3\cos\theta_k}\big)\\ -\frac{iut_1t_2\sin k}{4E_+^3\sin\theta_k}
			\end{pmatrix}dk\\
			&=i\int_{BZ}\left(-i\cos^2\theta_k\frac{\partial\phi_k}{\partial k}+\frac{iut_1t_2\sin k}{4E_+^3}-\frac{iut_1t_2\sin k}{4E_+^3}\right)dk\\
			&=\int_{BZ}\cos^2\theta_k\frac{\partial\phi_k}{\partial k}dk.
		\end{align*}
		Similarly, the expression for $Q^c_-$ can be obtained as,
		\begin{align*}
			Q^c_-&=i\int_{BZ}\begin{pmatrix}
				-e^{i\phi_k}\sin\theta_k\\ \cos\theta_k
			\end{pmatrix}^T\frac{\partial}{\partial k}\begin{pmatrix}
				-e^{-i\phi_k}\sin\theta_k\\ \cos\theta_k
			\end{pmatrix}dk\\
		&=\int_{BZ}\sin^2\theta_k\frac{\partial\phi_k}{\partial k}dk.
		\end{align*}
		Hence, the global Berry phase is given by,
		\begin{align*}
			Q^c_G&=\int_{BZ}\cos^2\theta_k\frac{\partial\phi_k}{\partial k}dk+\int_{BZ}\sin^2\theta_k\frac{\partial\phi_k}{\partial k}dk=\int_{BZ}d\phi_k
		\end{align*}
		Now, if this line integral over Brillouin Zone contains the origin, that is $t_1=t_2=0$, then it gives a value $2\pi$, otherwise it will give $0$.
		So, it follows that,
		\begin{equation*}
			Q^c_G=\begin{cases}
				2\pi\quad\mathrm{for}\quad t_1<t_2\\
				0\quad\mathrm{for}\quad t_1>t_2
			\end{cases}.
		\end{equation*}\vfill
	\end{appendices}

	\bibliographystyle{ieeetr}
	\bibliography{Final_draft}
\end{document}